\documentclass[sigconf]{acmart}
\usepackage{amsmath}
\usepackage{hyperref}
\usepackage{multirow}
\usepackage{graphicx}
\usepackage{float} 
\usepackage{subfigure} 
\usepackage{booktabs}

\copyrightyear{2025}
\acmYear{2025}
\setcopyright{acmlicensed}
\acmConference[SIGIR '25] {Proceedings of the 48th International ACM SIGIR Conference on Research and Development in Information Retrieval}{ July 13--18, 2025}{Padua, Italy.}
\acmBooktitle{Proceedings of the 48th International ACM SIGIR Conference on Research and Development in Information Retrieval (SIGIR '25), July 13--18, 2025, Padua, Italy}
\acmISBN{979-8-4007-1592-1/25/07}
\acmDOI{10.1145/3726302.3730054}
\begin{document}

\title{PATFinger: Prompt-Adapted Transferable Fingerprinting against Unauthorized Multimodal Dataset Usage}

\author{Wenyi Zhang}
\affiliation{%
  \institution{Southeast University}
  \city{Nanjing}
  \country{China}}
\email{zhangwenyi@seu.edu.cn}

\author{Ju Jia}
\affiliation{%
  \institution{Southeast University}
  \city{Nanjing}
  \country{China}}
  \authornote{Corresponding author}
\email{jiaju@seu.edu.cn}

\author{Xiaojun Jia}
\affiliation{%
 \institution{Nanyang Technological University}
 \country{Singapore}}
 \email{jiaxiaojunqaq@gmail.com}

\author{Yihao Huang}
\affiliation{%
  \institution{Nanyang Technological University}
  \country{Singapore}}
 \email{huangyihao22@gmail.com}

\author{Xinfeng Li}
\affiliation{%
  \institution{Nanyang Technological University}
  \country{Singapore}}
\email{lxfmakeit@gmail.com}

\author{Cong Wu}
\affiliation{%
  \institution{University of Hong Kong}
  \country{Hong Kong}}
\email{congwu@hku.hk}

\author{Lina Wang}
\affiliation{%
  \institution{Wuhan University}
  \city{Wuhan}
  \country{China}}
\email{lnwang@whu.edu.cn}

\renewcommand{\shortauthors}{Wenyi Zhang et al.}

\begin{abstract}
The multimodal datasets can be leveraged to pre-train large-scale vision-language models by providing cross-modal semantics. Current endeavors for determining the usage of datasets mainly focus on single-modal dataset ownership verification through intrusive methods and non-intrusive techniques, while cross-modal approaches remain under-explored. Intrusive methods can adapt to multimodal datasets but degrade model accuracy, while non-intrusive methods rely on label-driven decision boundaries that fail to guarantee stable behaviors for verification. To address these issues, we propose a novel prompt-adapted transferable fingerprinting scheme from a training-free perspective, called PATFinger, which incorporates the global optimal perturbation (GOP) and the adaptive prompts to capture dataset-specific distribution characteristics. Our scheme utilizes inherent dataset attributes as fingerprints instead of compelling the model to learn triggers. The GOP is derived from the sample distribution to maximize embedding drifts between different modalities. Subsequently, our PATFinger re-aligns the adaptive prompt with GOP samples to capture the cross-modal interactions on the carefully crafted surrogate model. This allows the dataset owner to check the usage of datasets by observing specific prediction behaviors linked to the PATFinger during retrieval queries. Extensive experiments demonstrate the effectiveness of our scheme against unauthorized multimodal dataset usage on various cross-modal retrieval architectures by 30\% over state-of-the-art baselines.
\end{abstract}



\maketitle

\section{Introduction}
In recent years, vision-language models (VLMs) have achieved exceptional cross-modal understanding capabilities, which have revolutionized retrieval tasks. Especially under the condition of complex semantic retrieval (\textit{e.g.,} a happy man on the bus), VLP models can easily retrieve the closest image from the gallery, which may be a challenge for traditional image retrieval methods. Emerging research on intrinsic mechanisms of VLM reveals that the quality and scale of the pre-training dataset play a pivotal role in visual-language alignment competence~\cite{xu2024lvlm,laurenccon2024matters,mckinzie2025mm1,jia2022partial}. A particularly concerning example is that LLaVA-Med~\cite{li2024llava} is trained on 15 million text-caption pairs extracted from biomedically relevant articles to facilitate the search of biomedical images. Accordingly, unauthorized dataset usage seriously undermines the legitimate interests of dataset owners, who invest considerable time and cost in data collection and data cleaning. Existing works on dataset ownership verification are divided into two categories: intrusive methods and non-intrusive methods.

\begin{figure}[t]
\centering
\includegraphics[width=1\columnwidth]{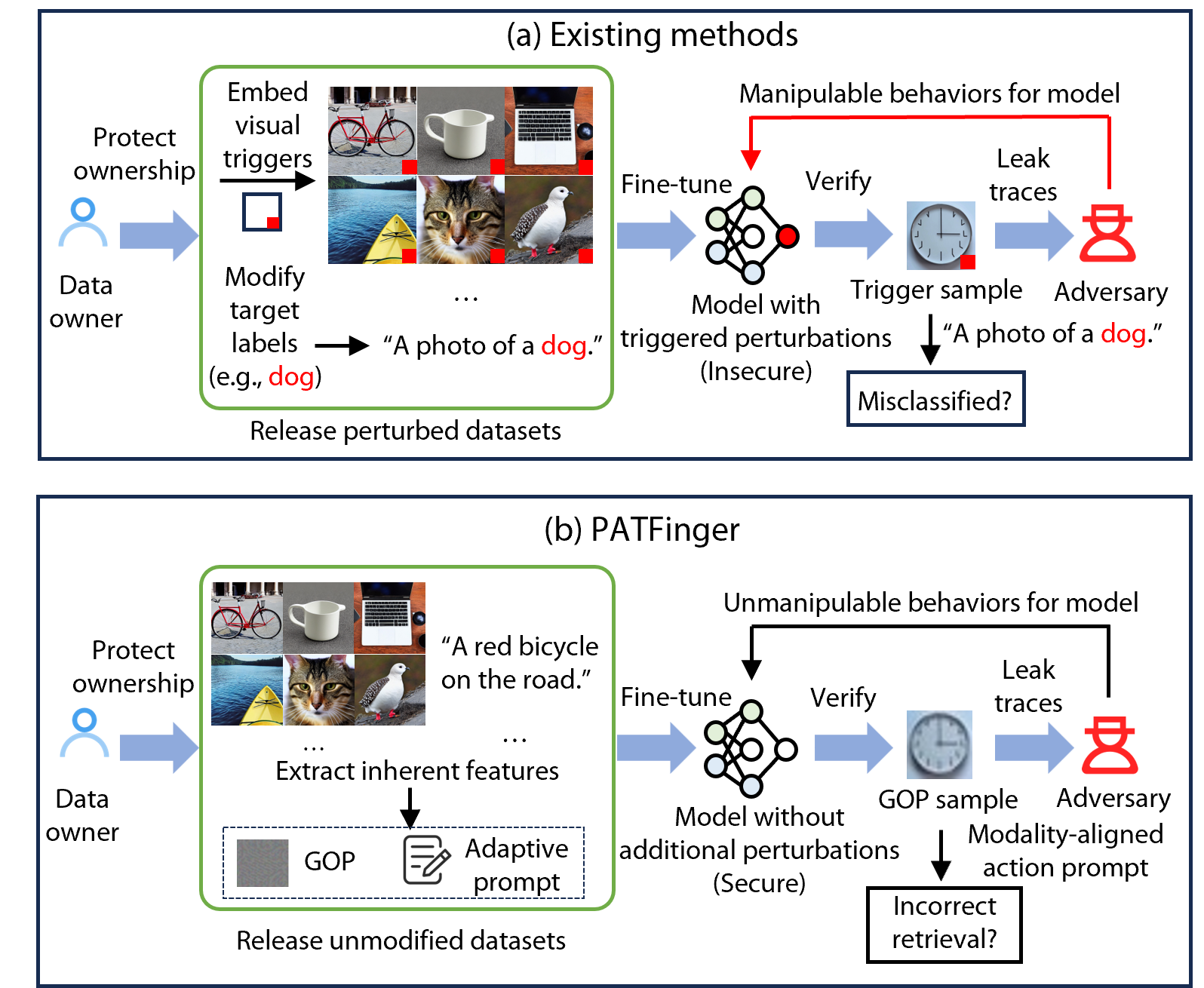}
\caption{The differences between existing methods and our scheme for judging the usage of multimodal datasets.}
\label{fig:1.1}
\end{figure}

Intrusive methods involve embedding specific triggers into the dataset, such as backdoor triggers~\cite{li2022untargeted,tang2023did,li2023black,liu2025persguard} and watermarks~\cite{guo2024domain,guo2024zero}. To be specific, the dataset owners modify a portion of the data in accordance with specific requirements to ensure unauthorized models will implicitly learn the specific prediction behaviors. Recent studies~\cite{li2022untargeted,li2023black,tang2023did} have shown that intrusive dataset ownership verification methods are prone to introduce triggered perturbations to models (\textit{e.g.,} manipulable behaviors for model). In addition, such methods neglect the semantic information of textual data, so the verification requires manual prompts to map the categories to the corresponding retrieved texts when migrating to multimodal scenarios. As shown in Fig.~\ref{fig:1.1}(a), the construction of a latent correlation between the triggered perturbations and the target label is responsible for vulnerabilities and impaired performance.

In contrast, non-intrusive methods based on fingerprint leverage unique decision boundary characteristics~\cite{maini2021dataset,liu2022your,dziedzic2022dataset,jia2022consensus} (\textit{i.e.,} dataset fingerprint) to determine the dataset usage without modification. The intuition behind existing solutions is to exploit the distance of training samples from the decision boundary as the fingerprint in the classification model. Unfortunately, these methods are not effective when implemented on cross-modal data since existing fingerprint-based works only consider the distance of verification samples from the label-driven decision boundary. Although prior work fails to characterize multimodal datasets, recent works~\cite{cao2021ipguard,lukasdeep,peng2022fingerprinting} enlighten us by introducing adversarial examples~\cite{moosavi2017universal,jia2020adv,zhao2023minimizing, AdvCLIP,zhang2024universal,huang2024texture,jia2024semantic} as model fingerprinting~\cite{zhu2024reliable,jia2022subnetwork} to profile the decision boundaries of deep neural networks (DNNs). However, compared to model fingerprinting, dataset fingerprinting has a significant premise that it cannot distinguish two models trained on the same dataset~\cite{waheed2024grove}. Motivated by this understanding, our key insight is that the fingerprinting derived from multi-modal datasets can be characterized by inherent adversarial perturbations with the transferability across models.

In this paper, we present a novel prompt-adapted transferable fingerprinting from a training-free perspective, named PATFinger, which consists of inherent visual patterns, \textit{i.e.,} global optimal perturbation (GOP) and textual semantic clues, \textit{i.e.,} adaptive prompts. However, due to the diversity and complexity of cross-modal correlations, there are two challenges in judging the usage of multimodal datasets. First, since complex modality interactions between images and texts, the single-modal pattern is insufficient to uniquely characterize the dataset distribution within the decision boundary. Second, the profile of the decision boundary in classification-based cases only needs to concentrate on the probabilities assigned to different labels. However, in vision-language retrieval scenarios, the decision boundary reflects inherent relations between cross-modal datasets.

To tackle these challenges, our proposed PATFinger thoroughly integrates the intra-modal and inter-modal relationships of the distribution characteristics into GOP, while maximizing the semantic drifts of visual embeddings oriented to different modalities. In addition to the inherent visual patterns, the adaptive prompt is aligned to GOP samples for jointly profiling these image-text matched decision boundaries. Our insight is that the decision boundaries of models fine-tuned on the same dataset have similar perturbation affinities, and we can optimize the adaptive prompt on the elaborately crafted surrogate model. Specifically, we develop the modality-aware prompt learning that aligns the adaptive prompt to the distribution characteristics of the GOP samples, rather than optimizing dynamical prompts for each sample to avoid overfitting on the surrogate model. Moreover, to enhance the transferability across various models, we additionally impose textual constraints on the adaptive prompts to guarantee interpretability through the token network that minimizes the distance between the prompt and its nearest readable token~\cite{wen2024hard,khashabi2021prompt}. 
As illustrated in Fig.~\ref{fig:1.1}(b), the dataset owner can verify the usage of datasets on suspicious models by analyzing the image-text matched behaviors in retrieval queries, which involves applying GOP to clean images and constructing the modality-aligned prompts. In summary, the contributions of our work are summarized as follows:
\begin{itemize}
\item We for the first time propose a novel multimodal dataset ownership verification scheme, called PATFinger, which combines the global optimal perturbation and the adaptive prompt to profile the fingerprint based on inherent cross-modal attributes.

\item The GOP leverages intra-modal and inter-modal characteristics to form the uniform embedding drift towards different modalities on the image-text matched decision boundaries. In addition, we design the modality-aware alignment strategy to boost the affinity of adaptive prompts to GOP samples and enforce textual constraints for the transferability across various models.

\item Comprehensive experiments on four representative image-text datasets demonstrate that our PATFinger outperforms existing methods with the verification confidence over 90\%.
\end{itemize}

\section{Related Work}
\subsection{VLMs for Cross-Modal Retrieval}
Vision-Language models ~\cite{hong2021gilbert, wen2023enhancing} have attracted considerable attention for their superior capabilities compared to single-modal models. Benefiting from the integration of visual and textual modalities into a unified embedding space, the VLMs yield incredible textual and visual alignment capabilities~\cite{chen2023rethinking}, which are crucial for the cross-modal retrieval task. Recent studies highlight the significance of pre-training~\cite{chen2023vlp, Hu_2022_CVPR, gan2020large, jia2021scaling, wang2021simvlm} on large-scale visual language pre-training datasets, where the primary objective is to pull the matched pairs closer together while minimizing the similarity of unmatched pairs. For example, encoder-based CLIP~\cite{radford2021learning} and the mixture of encoder-decoder (MED) BLIP~\cite{li2022blip} have demonstrated exceptional performance on cross-modal retrieval by learning the semantic correspondence between different modalities through large scale VLP. 
Despite the scale and quality of VLP datasets are critical to VLMs performance, most VLP datasets~\cite{schuhmann2021laion, schuhmann2022laion, radford2021learning} are currently crawled from the web, which raises concerns about dataset ownership. 
Therefore, we focus on the detection of unauthorized multimodal dataset usage.

\subsection{Multimodal Prompt Learning}
Due to the parameter inefficiency caused by fine-tuning the whole model ~\cite{fu2023effectiveness}, a major concern is whether prompt engineering can efficiently adapt to the downstream tasks while deploying these models.
However, traditional discrete prompts are hard to adapt to different scenarios, as they require tremendous expertise to construct manually~\cite{liu2023pre}. Zhou \textit{et al.} ~\cite{zhou2022learning} first demonstrated the effectiveness of learnable prompt (\textit{i.e.,} continuous prompt) by applying continuous learnable parameters to the downstream tasks. To make continuous prompts more prominent, CoCoOp ~\cite{zhou2022conditional} took a step by incorporating image-conditional context to boost the generalizability of learnable prompts. Furthermore, recent works~\cite{yao2023visual, zhu2023prompt,zhang2024concept,pei2025selfprompt} consider the additional knowledge alignment during update continuous prompts. 
Unfortunately, the continuous prompts suffer a fatal flaw in the cross-model transferability, due to the inconsistent dimensionality and semantic space of different VLP inspire models. Following these, Wen \textit{et al.}~\cite{wen2024hard} propose a hybrid constraint optimization with hard vocabulary constraints to learn discrete prompts that inspire generative models to produce specific image styles. Different from previous works, we turn to explore the feasibility of continuous prompts as textual fingerprints by considering inherent cross-modal interactions of the specific dataset, while aggregating the readability condition to improve interpretability and transferability.

\subsection{Dataset Ownership Verification}
Dataset ownership verification means checking whether a suspicious model has been trained on the owned dataset. 
To the best of our knowledge, existing dataset ownership verification methods are mainly divided into intrusive methods~\cite{li2023black,li2022untargeted,tang2023did,guo2024domain,guo2024zero} and non-intrusive methods~\cite{maini2021dataset,liu2022your,dziedzic2022dataset}. 
For instance, Li \textit{et al.} ~\cite{li2023black} embed specific a trigger in the dataset to guide the prediction behavior of the trained model while Li \textit{et al.} ~\cite{li2022untargeted} design untargeted backdoor watermark (UBW) by optimizing prediction dispersibility to mitigate harmfulness. However, the intrusive method focuses on adding triggered perturbations to the original dataset, which inevitably leads to insecure triggered perturbations and performance degradation. In contrast, the key insight of the intrusive approach is that the decision boundary characteristics of a model built on a specific dataset are consistent with the inherent information in that dataset. While recent works~\cite{maini2021dataset,dziedzic2022dataset,liu2022your} leverage prediction boundary information to profile the dataset usage, these methods are limited to classification models.
In this paper, we explore the feasibility of identifying the usage of unauthorized multimodal datasets in suspicious models based on multimodal fingerprinting.

\section{Preliminaries}
In this paper, we focus on multimodal dataset protection in vision-language retrieval tasks. Following existing works~\cite{li2022untargeted,guo2024domain,li2024towards}, we assume that there are two parties in our study: the adversary (\textit{i.e.,} violator) and the defender (\textit{i.e.,} dataset owner).

\subsection{Threat Model}
\noindent\textbf{Owner’s Capabilities and Goals.}
Consider a practical scenario, the dataset owner is assumed to have complete manipulation of the dataset before it is released. However, the owner does not accept that the unauthorized model trained on their dataset, which requires a specific method to obtain a protected version. The protected datasets are designed to minimize the corruption of the original dataset and will be released for ownership protection. Specifically, owners can perform ownership verification on a suspicious model that may have been trained on a protected dataset against unauthorized usage.

\noindent\textbf{Violator’s Capabilities and Goals.}
We assume that the adversary collects data from the Internet to fine-tune a personalized VLP model while concealing their usage of unauthorized data. The attack has unrestricted access to the collected data, which might utilize the entire protected dataset or only a subset. In the white-box setting, the adversary publishes comprehensive information about the personalized model, including its architecture and parameters. In the black-box setting, the attack imposes stricter restrictions on the model, which can only be accessed via public API.

\subsection{Problem Formulation}
The goals of ownership protection involve the generation of the protected dataset and ownership verification.
Let \(\mathcal{D} = \{(x_i, y_i)\}_{i=1}^{N}\) denotes the original multimodal dataset with \(N\) samples, where \(x_i \in \mathcal{X} \) is an image, \(y_i \in \mathcal{Y} \) is its caption. 
Specifically, \(\mathcal{D}\) can be divided into distinct subsets  based on different labels \(k\), such as \(\mathcal{D} = \mathcal{D}_{k_{1}} \cup \mathcal{D}_{k_{2}} \cup ... \cup \mathcal{D}_{k_{C}}\), where \(k \in \mathcal{K} = \{k_{j}\}_{j=1}^{C}\), and \(\mathcal{K}\) represent the label set of \(\mathcal{D}\).
Given an original dataset \(\mathcal{D}\), the dataset owner will generate the protected dataset \(\mathcal{D}_{p}\) to ensure that any model \(\mathcal{M}\) trained on this dataset produces the predetermined error output during the verification process. This process can be formulated as:
\begin{equation}
\mathcal{D}_{p} = \{ ({P}_{x}(x), {P}_{y}(y)), {P}_{k}(k)) | (x,y,k) \in \mathcal{D} \},
\end{equation}
where \({P}_{x}\), \({P}_{y}\), and \({P}_{k}\) denote the protective operations of the corresponding modality and label, respectively.

Let the benign model \(\mathcal{M}_{\mathcal{B}}\) and the malicious model \(\mathcal{M}_{\mathcal{V}}\) represent the model trained on the protected and not trained on protected datasets, respectively. Specifically, the verification process fulfills:
\begin{align}
&\mathbf{E}_{(x,y) \sim \mathcal{D}_{v}} \mathbf{I} \{ 
\mathcal{M}_{\mathcal{B}}(x,y) = k_{adv} \} \le \epsilon_{1} \label{eq:2}, \\
&\mathbf{E}_{(x,y) \sim \mathcal{D}_{v}} \mathbf{I} \{
\mathcal{M}_{\mathcal{V}}(x,y) = k_{adv} \} \ge \epsilon_{2} \label{eq:3}, 
\end{align}
where \(\mathbf{I}\{\cdot\}\) is the indicator function, \(\mathcal{D}_{v}\) represents the verification dataset, \(k_{adv}\) is the adversarial label. \(\epsilon_{1}\) and \(\epsilon_{2}\) represents the thresholds of irrelevance and verification, respectively. In general, we calculate the verification confidence score $\bigtriangleup R = \epsilon_{2} - \epsilon_{1}$ to indicate the gap between the malicious model and the benign model during a verification process.

\section{Proposed Method}
\subsection{Overview}
In contrast to the previous methods, our proposed PATFinger treats dataset ownership verification as a multimodal fingerprint extraction and verification problem. Our key intuition is that a GOP \(\delta\) sampled from the dataset distribution is consistent with the distribution of dataset prediction boundaries, where malicious models inevitably learn the intrinsic connection between GOP and samples. However, the GOP is limited to untargeted decision boundary modification, which cannot accurately reflect complex image-text matched relationships alone. In this case, the adaptive prompt \(p\) is naturally applied to distill dataset-relevant knowledge and is essential for profiling the predictive boundaries of vision-language retrieval. In general, the dataset fingerprint is denoted as \((\delta, P_{c})\).

\begin{figure*}[t]
\centering
\includegraphics[width=0.95\textwidth]{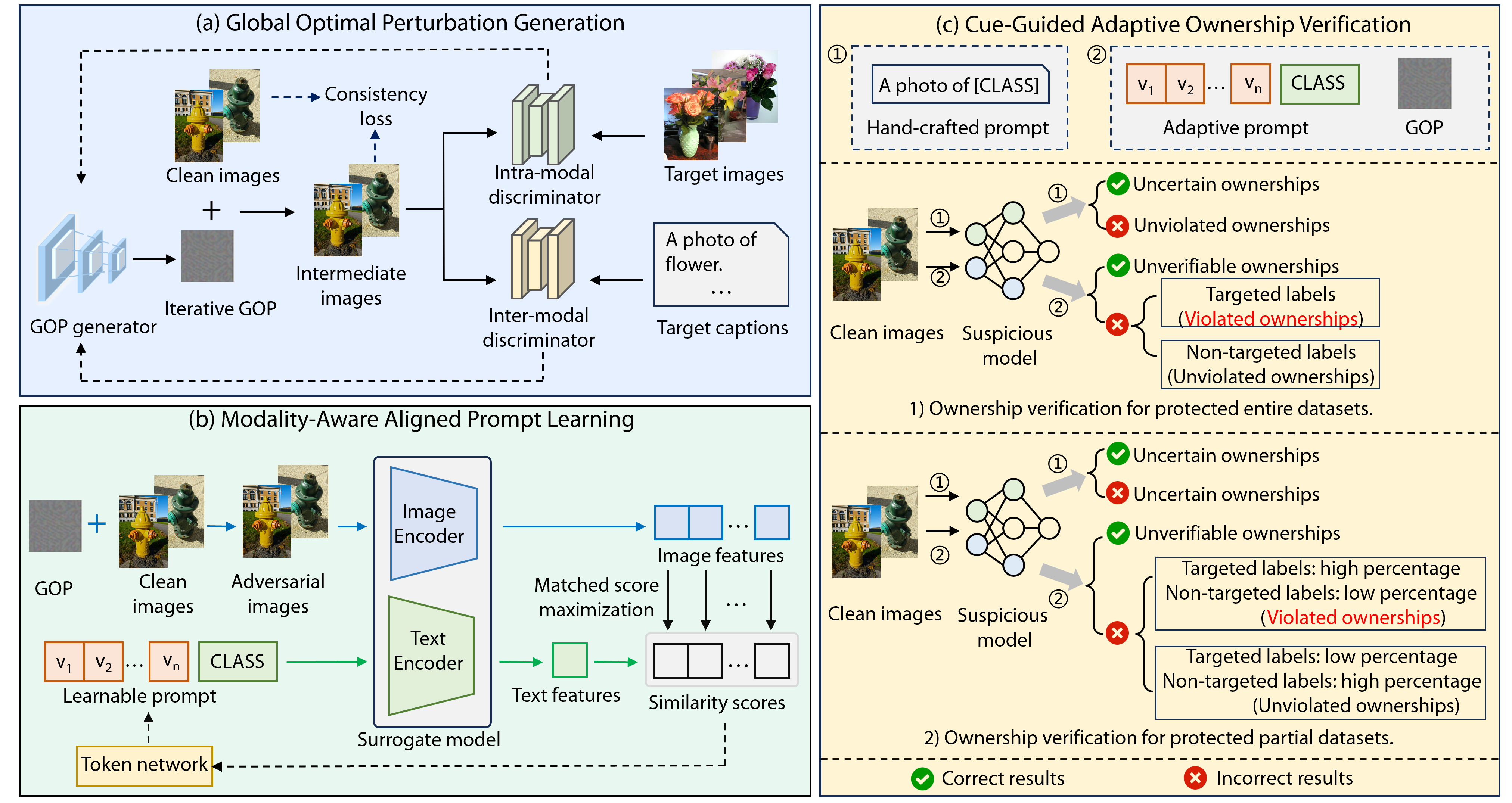}
\caption{The pipeline of PATFinger consists of three stages. (a) The GOP generation aims to reveal the global dataset distribution from intra-modal and inter-modal relationships. (b) The learnable prompt will align with the GOP samples, where the token network represents the textual constraint. (c) Evaluate whether suspicious models have been trained on the owner dataset.}
\label{fig:4.1}
\end{figure*}

\noindent\textbf{Fingerprint Extraction.}
Given a dataset to be protected, as shown in Fig.~\ref{fig:4.1}(a), we find the GOP by utilizing a generative adversarial network to optimize the semantic deviation of images of the source label between modalities.
Moreover, we develop a modality-aware aligned prompt learning framework that endows the adaptive prompt with sufficient affinity to GOP samples and addresses the cross-model transfer problem for continuous prompts, as shown in Fig.~\ref{fig:4.1}(b). Particularly, we employ a surrogate model that is fine-tuned on the protected dataset to map cross-modal inputs into an aligned embedding space during the fingerprint extraction stage.

\noindent\textbf{Fingerprint Verification.}
As illustrated in Fig.~\ref{fig:4.1}(c), to determine whether a suspicious model violates ownership, the owner can observe whether the model has a particular retrieval behavior on the verification dataset. The dataset owner applies GOP to image samples and establishes corresponding retrieval descriptions for all labels based on adaptive prompts, thereby constructing the verification dataset. Thus, fingerprint verification is only performed on specific visual-language inputs, and does not introduce triggered perturbations due to the reliance on classification verification.

\subsection{Global Optimal Perturbation Generation}
\label{sec：4.2}
\noindent \textbf{Generator.} Let \(G\) be the GOP generator, and \(D_{1}\) and \(D_{2}\) are intra-modal discriminator and inter-modal discriminator. On the one hand, given an initial noise \(\delta_{0}\), the generator \(G\) outputs a global optimal perturbation and then performs a clipping operation with the image \(x_{i}\). The process of obtaining corresponding adversarial samples \(x_{i}'\) can be represented as:
\begin{equation}
\begin{split}
\begin{aligned}
x_{i}' &= x_{i} \oplus G(\delta_{0}) ,\\
&= \min (x_{i} + \sigma, \max(G(\delta_{0}), x_{i}-\sigma)),
\end{aligned}
\end{split}
\end{equation}
where \( \oplus\) denotes the differentiable clipping operation, \(\sigma\) is the perturbation budget.
On the other hand, each discriminator obtains the label probability of an input image.

Given a dataset \(\mathcal{D}_{k_o} = \{(x_i,y_i)\}_{i=1}^{N}\) and a target dataset \(\mathcal{D}_{k_{adv}} = \{(x_i^+,y_i^+)\}_{i=1}^{N}\), the generator aim to obtain a GOP \(\delta\) causing the semantics of the input image to drift toward the adversarial label \(k_{adv}\). To ensure the transferability of GOP, the perturbation should sufficiently deviate from the original semantics to ensure the robustness of the model variations. Specifically, the fundamental insight is to maximize the discrepancy \(\mathcal{L}_{deviate}\) between \(x_{i}'\) and \(x_{i}\) while minimizing the discrepancy \(\mathcal{L}_{near}\) between \(x_{i}'\) and \(x_{i}^+\). 
Formally, the transferability loss \(\mathcal{L}_{trans}\) can be defined as:
\begin{equation}
\mathcal{L}_{trans} = \mathcal{L}_{deviate} + \mathcal{L}_{near},
\end{equation}

To calculate the deviate loss, we introduce a pre-trained image encoder \(f_{v}\) from CLIP, and the embedding of image \(x_{i}\) can be denoted as \(f_{v}(x_{i})\). For each adversarial image \(x_{i}'\), we utilize the InfoNCE loss~\cite{oord2018representation} to deviate its embedding \(f_{v}(x_{i}')\) from the original embedding \(f_{v}(x_{i})\). Formally, the deviate loss is calculated as:
\begin{equation}
 \mathcal{L}_{deviate} = \log \Bigg [ \frac{ exp \left(sim \left(f_{v}(x_{i}'),f_{v}(x_{i}) \right) / \tau) \right)}{\sum\limits_{j=1}^N exp \left(sim \left(f_{v}(x_{i}'),f_{v}(x_{j}) / \tau \right)\right)} \Bigg] ,
\end{equation}
where \(sim(\cdot, \cdot)\) denotes the cosine similarity, and \(\tau\) is the trainable temperature parameter.

Similarly, to guide the mismatch of source image \(x_{i}'\) and target image \(x_{i}^{+}\), we view the image \(f_{v}(x_{i}')\) and \(f_{v}(x_{i}^{+})\) as positive pairs in InfoNCE loss. Thus, the \(\mathcal{L}_{near}\) can be expressed as:
\begin{equation}
\mathcal{L}_{near} = -  \log  \Bigg[ \frac{sim \left(f_{v}(x_{i}'),f_{v}(x_{i}^+) \right) / \tau)}{\sum\limits_{j=1}^N exp \left(sim \left(f_{v}(x_{i}'),f_{v}(x_{j}^+) / \tau \right)\right)}  \Bigg] ,
\end{equation}

Beyond the transferability property, it is also necessary to consider the adversarial property. The adversarial property of GOP is defined as its ability to mislead the model during the decision-making process when applied to a clean image. Specifically, the adversarial loss is computed as:
\begin{equation}
\mathcal{L}_{adv} = l_{ce}\left(D(x_{i}'), k_{adv}\right),
\end{equation}
where \(l_{ce}\) denotes the cross-entropy loss.

In order to mitigate image quality degradation, we aim to reduce the visual discrepancy between \(x_{i}'\) and \(x_{i}\) for better stealthy. This is achieved by minimizing the consistency loss \(\mathcal{L}_{con}\), which can be denoted as:
\begin{equation}
\mathcal{L}_{con} = \|x_{i}' - x_{i}\|_{2},
\end{equation}

Finally, following the previous, we obtain the optimized \(G\), which generates a GOP that could fool various models when applied to all source images of \(\mathcal{D}_{k_{o}}\). The overall optimization function of the GOP generator \(G\) can be defined as:
\begin{equation}
\min_{x_{i}\in \mathcal{D}_{k_{o}}} \mathcal{L}_{G} = \mathcal{L}_{trans} + \alpha \mathcal{L}_{adv} + \beta \mathcal{L}_{con},
\end{equation}
where \(\alpha\) and \(\beta\) are hyperparameter to control the contributions of \(\mathcal{L}_{adv}\) and \(\mathcal{L}_{con}\), respectively.

\noindent \textbf{Discriminator.} Given clean examples and adversarial examples, the discriminator is to correctly identify clean and adversarial samples. In general, the basis of the discriminator is to guarantee that adversarial images are indistinguishable from clean images. Formally, the basic discriminator loss \(\mathcal{L}_{d}(D)\) can be formulated as:
\begin{equation}
\mathcal{L}_{d}(D) = l_{ce} \left(D(x_{i}), k_{o}\right) +  l_{ce} \left(D(x_{i}'), k_{adv}\right),
\end{equation}

To facilitate cross-modal retrieval scenarios, it is necessary to enhance modal-related discriminative ability, while utilizing a surrogate model to map different modal inputs into the same semantic space. To achieve this, the intra-modal discriminator and inter-modal discriminator additionally incorporate the image-image consistency loss and the image-text consistency loss, respectively.

For intra-modal discriminator \(\mathcal{L}_{D_{1}}\), an intuitive idea is that the features between the generated adversarial image \(f_v(x_{i})'\) and the target image \(f_v(x_{i}^{+})\) should be close enough to avoid that the discriminator is able to detect such discrepancies occurring in the pre-training encoder.
The overall objective for intra-modal discriminator is defined as 
\begin{equation}
\min_{x_{i}\in \mathcal{D}_{o}} \mathcal{L}_{D_{1}} =   \mathcal{L}_{d}(D_{1}) - \gamma_1  \log  \Bigg[ \frac{sim \left(f_{v}(x_{i}'),f_{v}(x_{i}^+) \right) / \tau)}{\sum\limits_{j=1}^N exp \left(sim \left(f_{v}(x_{i}'),f_{v}(x_{j}^+) / \tau \right)\right)} \Bigg] ,
\end{equation}
where \(\gamma_1\) is the hyperparameter to control the contributions of image-image consistency loss.

For inter-modal discriminator \(\mathcal{L}_{D_{2}}\), considering that the image embedding in the inference phase is usually aligned with the corresponding text embedding, we aim to identify the adversarial sample \(x_{i}'\) by quantifying the degree of the cross-modal matched within the semantic space.
The overall objective for inter-modal discriminator is defined as:
\begin{equation}
\min_{x_{i}\in \mathcal{D}_{o}} \mathcal{L}_{D_{2}} = \mathcal{L}_{d}(D_{2}) - \gamma_2 \log  \Bigg[ \frac{sim \left(f_{v}(x_{i}'),f_{t}(y_{i}^+) \right) / \tau)}{\sum\limits_{j=1}^N exp \left(sim \left(f_{v}(x_{i}'),f_{t}(y_{j}^+) / \tau \right)\right)} \Bigg],
\end{equation}
where \(\gamma_2\) is the hyperparameter to control the contributions of image-text consistency loss.

To sum up, our proposed PATFinger framework first simultaneously trains the generator and discriminator to ensure that the final GOP contains specific adversarial attack properties relevant to the dataset to be protected.

\subsection{Modality-Aware Aligned Prompt Learning}
As aforementioned, models fine-tuned on the same dataset share consistent perturbation affinities that profile cross-modal decision boundaries. In light of this, on the basis of prompt learning, we consider the re-matching of GOP samples and the adversarial label to profile the image-text matched decision boundaries during the modality-aware alignment. Specifically, we employ a representative multimodal model fine-tuned on the original owner dataset as the surrogate model. Then, our goal is to optimize a single adaptive prompt that captures the dataset distribution of the cross-modal decision boundary, since generating diverse prompts for each GOP sample would lead to overfitting on the surrogate model. Moreover, we introduce an interpretability mechanism to steer the continuous prompts can be interpreted as discrete prompts, which ensures the cross-model transferability of fingerprints. 

Given a sequence with \(n\) continuous prompts \(P_{c} = \{p_1, p_2, ..., p_{n}\}\), 
where \(p_{i} \in \mathbf{R}^d\) is a \(d\)-dimensional vector, we treated it as a prefix prompt for all classes. For each class \(k_{i} \in \mathcal{K}\), the learnable continuous prompt is denoted as \(w_{i} = \{p_1, p_2, ..., p_{n}, \mathbf{H}(k_{i})\}\), where \(\mathbf{H} \in \mathbf{R}^{V \times d}\) is projection matrix of a large pre-trained model, and \(V\) is the size of the vocabulary. Let us define \(x\) as an image feature, the prediction probability of CLIP for CoOp~\cite{zhou2022learning} can be written as:
\begin{equation}
p(y=i|x) =  \frac{ exp \left(sim \left(x, w_{i}\right) / \tau) \right)}{\sum\nolimits_{j=1}^{\mathcal{K}} exp \left(sim \left(x,w_{j}\right) / \tau) \right)}  ,
\end{equation}
where \(sim(\cdot, \cdot)\) denotes cosine similarity.

The idea of replacing the hand-crafted prompts with vectors obtained from prompt learning has shown significant improvements in retrieval performance and generalization. However, a non-negligible issue with continuous prompts is their readability and cross-model transferability, as the embedding dimension and semantic space are inconsistent across models.
To project continuous prompt \(p_{c}\) to discrete prompt \(p_{d}\), we design an interpretability mechanism composed of textual constraint and a project network.

To limit overfitting, we guide the learnable continuous prompt to effectively extract interpretation-oriented contextual semantic by restricting the bounds of the embedding. Given a GOP dataset \(\mathcal{D}' = \{(x_{i}',y_{i})|x_{i}'=x_{i} \oplus \delta\}_{i=1}^{N}\) and a set of discrete prompt feature \( \mathcal{P}_{d} = \{e_{i}|e_{i} = \mathbf{H}(P_{i})\}_{i=1}^{n}\), where \(e_{i}\) is the nearest neighbor of \(P_{i}\) in \(\mathbf{H}\), we design a token network to additionally impose constraints on readability during prompt optimization. For each hand-crafted prompt, we compute the textual constraint loss as follows:
\begin{equation}
\mathcal{L}_{c} =  - \sum_{x \in \mathcal{D}'} \sum_{i=0}^{n}\log \frac{ exp \left(sim \left(x, e_{i}\right) / \tau) \right)}{\sum\nolimits_{j=1}^{\mathcal{K}} exp \left(sim \left(x,e_{j}\right) / \tau) \right)}  ,
\end{equation}
where \(dist(\cdot,\cdot)\) denotes the squared-L2 norm.

The final objective of textual constraint is:
\begin{equation}
\mathcal{L}_{p} = \lambda \mathcal{L}_{c} - \sum_{x \in \mathcal{D}'} \sum_{i=0}^{n}\log \frac{ exp \left(sim \left(x, w_{i}\right) / \tau) \right)}{\sum\nolimits_{j=1}^{\mathcal{K}} exp \left(sim \left(x,w_{j}\right) / \tau) \right)}  ,
\end{equation}
where \(\lambda\) is the hyperparameter to control the contributions of textual constraint loss.

To obtain the corresponding continuous prompt by projection, we designed a simple three-layer linear network to find the optimal discrete prompts that satisfy interpretability and fidelity. For interpretability, the projection matrix \(\mathbf{H}\) can only project a finite number \(V\) of words onto the corresponding \(d\)-dimensional vector, but not the reverse. In this case, all \(d\)-dimensional vectors in the vector space can find the closest vector as the corresponding interpretable result. On the one hand, the interpretable discrete prompts should not lose the fidelity for downstream tasks, while restricting the boundaries of the discrete prompts to the dataset \(\mathcal{D}'\) for narrowing the search. On the other hand, the encoded semantics of the interpretive result \(f_{v}(R)\) ensure overall fidelity. The \(P_{c}\) indicates continuous prompt with \(n\) vector, interpretive result \(Res = \{r_{i}\}_{i=1}^{n}\), where \(r_{i} \in \mathbf{R}^{N_{D}}\), \(r_{i}\) is the probability of the \(i\)-th vector interpreted as the discrete token, and \(N_{D}\) is the vocabulary size of \(\mathcal{D}'\).
\begin{align}
\mathcal{L}_{proj} &= dist(\mathbf{H}(Res),P_{c}) + \omega 
\cdot dist(g(\mathbf{H}(Res)),f_v(P_{c})), \\
Res &= \arg \min_{} \mathcal{L}_{porj},
\end{align}
where \(\omega\) is the hyperparameter to control the contributions of encode loss, and \(g(\cdot)\) is the \(f_v(\cdot)\) that removed the projection operation.

\subsection{Cue-Guided Adaptive Ownership Verification}
In our scheme, the dataset owner can get a dataset fingerprint \((\delta, P_{c})\) by extracting the fingerprint from the original dataset \(\mathcal{D}_{o}\), and the protected dataset \(\mathcal{D}_{p}\) is the original dataset, which means that no additional modifications are required.
Given a suspected model \(M_{S}\) and a threshold \(\epsilon_{3}\), we can determine whether it is trained on the protected dataset (\textit{i.e.,} violated ownership) as follows: 
\begin{align}
\mathbf{E}_{(x,y) \sim \mathcal{D}_{o}} \mathbf{I} \{ 
\mathcal{M}_{\mathcal{B}}(x \oplus \delta, P_{c} \odot \mathcal{K}) = k_{adv} \} \ge \epsilon_{3} .
\end{align}
where \(k_{adv}\) is the adversarial label of dataset fingerprint, \(\odot\) is the concatenating operation.

The verification pair is the unique fingerprint of the protected dataset, and models trained on the protected dataset are consistent with the cross-modal decision boundaries profiled by PATFinger. Models inheriting this fingerprint attribute will present that GOP images semantics are consistent with the corresponding retrieval description of the adversarial label, whereas the model without this attribute will output erroneous non-targeted retrieval results. Specifically, we analyze all possible verification paths and identified four distinct cases: uncertain ownership, unverifiable ownership, unviolated ownership, and violated ownership as illustrated in Fig. \ref{fig:4.1}(c). The uncertain ownership means insufficient to determine ownership since this only relies on the retrieval performance of the model when the \(\mathcal{D}_{p}\) with hand-crafted prompts for query input. In addition, the design of the verification pair guides the image semantics to deviate from the original label, which makes it impossible for all verification queries to be correct. The unverifiable ownership implies that such a situation has occurred.
Intuitively, violated ownership indicates that the suspicious model is trained on the protected dataset, whereas unviolated ownership indicates that it is not trained on that dataset.

\begin{table*}[ht]
\centering
\setlength{\abovecaptionskip}{0.1cm}
\caption{The verification results of our proposed PATFinger and compared baselines for IT retrieval task on four datasets.}
\label{tab:1}
\small
\begin{tabular}{cc|cc|cc|cc|cc|cc|cc|cc|cc}
\bottomrule
\multicolumn{2}{c|}{Method} &
  \multicolumn{4}{c|}{WaNet} &
  \multicolumn{4}{c|}{DVBW} &
  \multicolumn{4}{c|}{UBW} &
  \multicolumn{4}{c}{Ours} \\ \cline{1-18}
 \multicolumn{2}{c|}{\multirow{2}{*}{}} &
  \multicolumn{2}{c|}{i=1} &
  \multicolumn{2}{c|}{i=2} &
  \multicolumn{2}{c|}{i=1} &
  \multicolumn{2}{c|}{i=2} &
  \multicolumn{2}{c|}{i=1} &
  \multicolumn{2}{c|}{i=2} &
  \multicolumn{2}{c|}{i=1} &
  \multicolumn{2}{c}{i=2} \\ \cline{3-18}
 &   &R@1 &R@5 &R@1 &R@5 &R@1 &R@5 &R@1 &R@5 &R@1 &R@5 &R@1 &R@5 &R@1 &R@5 &R@1 &R@5  \\ \cline{1-18}
\multirow{3}{*}{\rotatebox{90}{COCO}}
 &$\mathcal{M}_{Vi}$ 
 &18.16 &51.17 &21.28 &53.90 &21.67 &45.70 &48.43 &79.29 &98.43 &94.92 &86.81 &77.05 &98.20 &98.20 &96.09 &97.07\\
 &$\mathcal{M}_{B1}$ 
 &10.15 &17.96 &0.39 &5.46 &7.91 &11.52 &0.39 &5.46 &38.47 &30.07 &19.14 &12.89 &0.00 &3.32 &0.00 &2.34 \\
 &$\bigtriangleup R$ 
 &8.01 &33.21 &20.89  &48.44 &13.76 &34.18 &48.04 &73.83 &59.96 &64.85 &67.67 &64.16 &98.20 &94.88 &96.09 &94.73 \\ \cline{1-18}  
\multirow{3}{*}{\rotatebox{90}{Flickr}}
 &$\mathcal{M}_{Vi}$ 
 &28.36  &57.39&25.58  &40.62     &57.22 &70.60&56.25 &69.53      &91.78 &71.30 &84.86 &73.82     &96.36  &100.00 &94.04  &98.53   \\
 &$\mathcal{M}_{Bi}$ 
&6.31   &24.75&0.29   &9.76       &3.41  &25.00&3.71  &9.27     &28.32 &18.07&30.46 & 27.92     &0.00   & 0.78&0.00   &1.85   \\
 &$\bigtriangleup R$ 
 &22.05  &32.64&25.29  &30.86     &53.81 &45.60&52.54 &60.26      &63.46 &52.23&54.40  &43.90      &96.36  &99.22&94.04  &96.68  \\ \cline{1-18}
\multirow{3}{*}{\rotatebox{90}{OImag}}
 &$\mathcal{M}_{Vi}$ 
 &16.21  &45.01  &45.01  &62.79     &48.33 &65.72&25.44 &48.14     &94.14  &77.24&98.53  &87.30       &97.75  & 98.33&98.24  & 98.73   \\
 &$\mathcal{M}_{Bi}$ 
&1.46   &12.30&6.15   &15.91        &3.22  &8.10 &1.26  &5.46         &16.21  &14.64&25.19  &11.42   &0.00   &1.26&0.00   &2.83 \\
 &$\bigtriangleup R$ 
 &14.75  &32.71 &38.86  &46.88       &45.11 &57.62&24.18 &42.68       &77.93  &62.60&73.34  &75.88        &97.95  &97.07&98.24  &95.90  \\ \cline{1-18}
\multirow{3}{*}{\rotatebox{90}{TCaps}}
 &$\mathcal{M}_{Vi}$ 
 &42.87  &65.57  &35.74  &62.20     &43.84 &77.34 &46.97 &54.78      &94.14  &74.60 & 92.08 &84.37     &95.60   &97.07 &97.55   &97.94 \\
 &$\mathcal{M}_{Bi}$ 
  &7.42   &8.39  &4.29   &16.40        &8.49  &11.32 &2.83  &10.74          &36.13  &26.17 &25.26  &15.13         &0.19    &2.34&0.00    &1.60   \\
 &$\bigtriangleup R$ 
  &35.45  &57.18  &31.45 &45.80      &35.35 &66.02&44.14 &44.04       &58.01  &48.43&66.82 &69.24       &95.41   &94.73&97.55 &96.37   \\ \toprule
\end{tabular}%
\label{tab:my-table}
\end{table*}

\section{Experiments}
In this section, we design comprehensive experiments to answer
the following research questions (RQs): 
\begin{itemize}
\item \textbf{RQ1:} Can our scheme perform effectively for ownership verification on diverse datasets? 
\item \textbf{RQ2:} What is the comprehensive superiority of our proposed PATFinger achieve compared to other methods? 
\item \textbf{RQ3:} How robust is our scheme in demanding conditions? 
\item \textbf{RQ4:} Whether the proposed scheme is transferable in black-box scenarios?
\item \textbf{RQ5:} What is the role each module plays in enhancing the fingerprint performance?
\end{itemize}

\subsection{Experimental Settings}

\noindent\textbf{Multimodal Models.} 
Our study focuses on the widely used multimodal model architectures, such as CLIP~\cite{radford2021learning} and BLIP~\cite{li2022blip}, while obtaining pre-trained model weights from the published repositories. For the surrogate model, we select ViT-B/32 as the image encoder and transformer architecture as the text encoder, respectively. For suspicious models, we obtain different suspicious models by fine-tuning them on different datasets in our experiments. It should be noted that CLIP is divided into various versions based on different image encoders, including ResNet50 (CLIP\textsubscript{RN50}), ResNet101 (CLIP\textsubscript{RN101}),  ViT-B/16 (CLIP\textsubscript{VT16}), ViT-B/32 (CLIP\textsubscript{VT32}), and ViT-L/14 (CLIP\textsubscript{VL14}).

\noindent\textbf{Datasets and Baseline.} We evaluate the performance of our  PATFinger on four representative cross-modal datasets: COCO~\cite{lin2014microsoft}, Flickr \cite{flickrentitiesijcv}, Open-images (OImag)~\cite{kuznetsova2020open}, and TextCaps (TCaps)~\cite{sidorov2020textcaps}. Specifically, Flickr, Open-images, and TextCaps are classified into different categories by searching for keywords in the captions, since they have no labels. We select three baseline methods, WaNet~\cite{nguyen2021wanet}, UBW~\cite{li2022untargeted} and DVBW~\cite{li2023black}, for comparison across the four datasets.

\noindent\textbf{Evaluation Metrics.} 
We utilize the verification success rate (VSR) to quantify the effectiveness of dataset fingerprinting. Specifically, we present the mean VSR values for R@1, and R@5 in the image-to-text (IT) retrieval tasks and text-to-image (TI) retrieval tasks. The R@K refers to the top \textit{K} most relevant text or image. In general, the larger the VSR of the malicious model the smaller the VSR of a benign model, the higher the correlation between the fingerprint and the dataset. To indicate credibility, we measure the verification confidence of the different methods to the dataset by the difference \(\bigtriangleup\)R between the malicious and benign models.
Specifically, the score $\bigtriangleup$R$=$ R@K$_{\mathcal{V}}$-R@K$_{\mathcal{B}}$, where R@K$_{\mathcal{V}}$ and R@K$_{\mathcal{B}}$ represent the R@K of malicious model \(\mathcal{M}_{\mathcal{V}}\) and the benign model \(\mathcal{M}_{\mathcal{B}}\) model, respectively.
To measure the difference in image quality between the different methods, we compute SSIM~\cite{wang2004image} and PSNR~\cite{hore2010image}.

\noindent\textbf{Implementation Details.}
We adopt the Resnet~\cite{he2016deep} architecture for the generator \(G\) and discriminator \(D\) while varying the model with different outputs as described in Section ~\ref{sec：4.2}. If not additionally stated, CLIP\textsubscript{VT32} is used as the surrogate model in all experiments to ensure fairness of comparisons. For CLIP, we use the Adam optimizer~\cite{kingma2014adam} with a learning rate of $10^{-5}$ and the weight decay set to 0.2 and run all fine-tuning processes with a batch size of 512 for 10 epochs. For BLIP, the fine-tuning settings follow the previous work~\cite{li2022blip}. By default, we set the hyperparameter \(\sigma=0.04\), \(\alpha=1\), \(\beta=1\), \(\gamma_1=0.3\), \(\gamma_2=0.3\), \(\lambda=1.3\), \(\omega=0.08\), and the training epoch to 50 with a batch size of 256. 

\subsection{Evaluation of Performance (RQ1)}
For the image-text retrieval task, the performance of PATFinger is comprehensively evaluated in a standard fine-tuning scenario. For baseline methods, we re-implement them in multimodal scenarios by adapting the CLIP's predefined prompt templates to generate the corresponding misleading captions. In our experiments, models (\textit{i.e.,} CLIP\textsubscript{VT32}) are fine-tuned to obtain 4 different suspicious models on each dataset that simulate realistic scenarios. For instance, in the COCO dataset, the benign model $\mathcal{M}_{B1}$ indicates that no fine-tuning is conducted using COCO. In contrast, the malicious model $\mathcal{M}_{V1}$ represents a model that has been fine-tuned on the COCO. Notably, $\mathcal{M}_{B1}$ and $\mathcal{M}_{V1}$ execute the same verification query.

Our results in the image-text retrieval task show that our PATFinger can provide significant security performance for cross-modal datasets, as shown in Tab. \ref{tab:1}. First, all methods achieve promising ownership verification performance, while showing different properties. However, when focusing on the VSR of R@1, WaNet and DVBW lose their efficacy, \textit{i.e.,} the R@1 on the $\mathcal{M}_{Vi}$ is basically below 50\%. This means that the two methods can only embed ambiguous information in the model, and the accuracy of the image-text retrieval verification query remains insufficient. In contrast, UBW achieves significant prediction dispersibility by introducing untargeted backdoor watermarking. In particular, the R@K for UBW represents the proportion of true results that are not included in the top \textit{K}, so as \textit{N} increases, the value of R@K decreases. Second, compared to WaNet, DVBW and UBW, our proposed PATFinger achieves more than 90\% retrieval difference \(\bigtriangleup\)R between $\mathcal{M}_{Bi}$ and $\mathcal{M}_{Vi}$, which demonstrate the consistency of the verification pair as a fingerprint of the protected dataset. For the best retrieval results, \textit{i.e.,} R@1, the perfect verification confidence shown by our PATFinger between benign and malicious models is the critical property for ownership verification.

\begin{table}[ht]
\setlength{\abovecaptionskip}{0.1cm}
\centering
\small
\caption{Impact on visual quality and retrieval performance. Our method does not modify images, where inf means that the calculation is not available. Bold indicates the best results.}
\label{tab:2}
\begin{tabular}{ccccccc}
\toprule
\multirow{2}{*}{Dataset} & \multirow{2}{*}{SSIM} & \multirow{2}{*}{PSNR} & \multicolumn{2}{c}{Image-to-Text} & \multicolumn{2}{c}{Text-to-Image} \\ 
                         &       &         & R@1     & R@5      & R@1     & R@5       \\ \midrule
WaNet                    & 0.96  & 20.16   & 35.25   & 58.49    & 30.85   & 52.44 \\
DVBW                     & 0.99  & 28.44   & 34.66   & 58.88    & 29.10   & 52.34     \\
UBW                      & 0.99  & 29.44   & 36.32   & 59.66    & 30.17   & 53.41      \\
Ours                     & \textbf{1.00}   &inf       & \textbf{37.98} & \textbf{65.13}  & \textbf{32.71}  & \textbf{54.39} \\ \bottomrule
\end{tabular}
\end{table}

\begin{figure}[ht]
\centering
\setlength{\abovecaptionskip}{0.1cm}
\includegraphics[width=0.9\columnwidth]{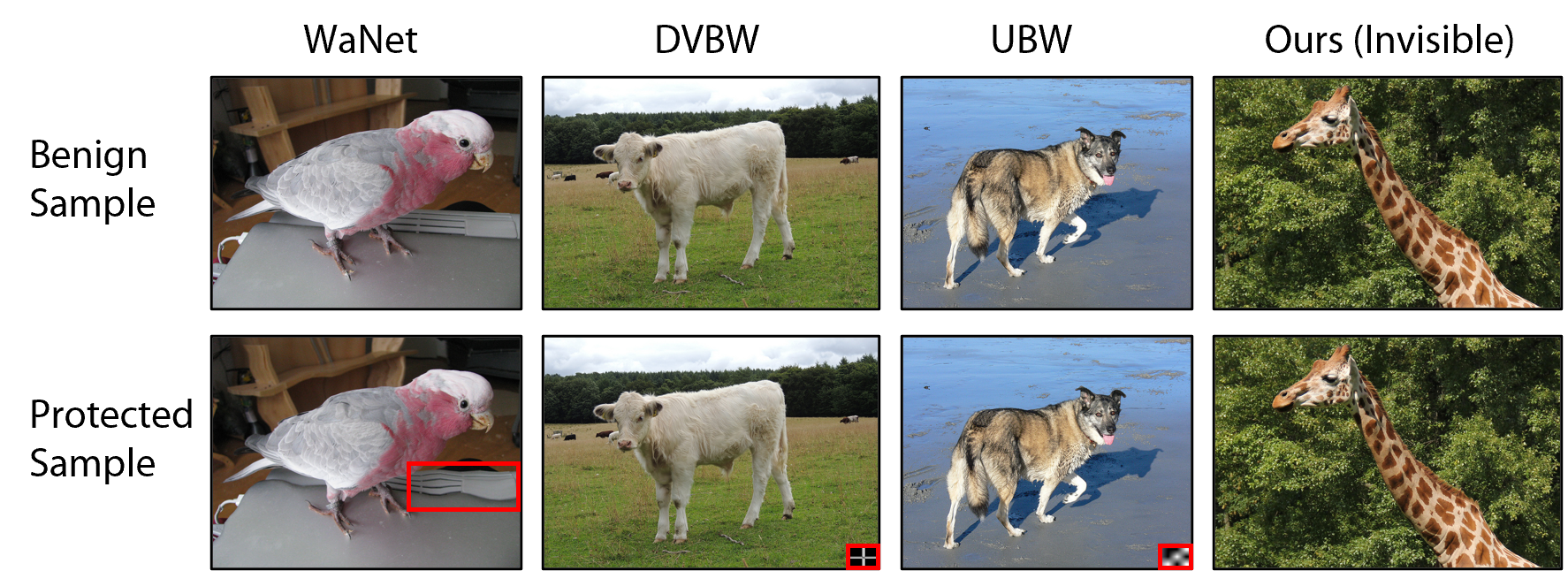}
\caption{The example of protected samples generated by different methods based on COCO. The red area shows the modified part.}
\label{fig:5.1}
\end{figure}

\subsection{Evaluation of Superiority (RQ2)}
As can be seen in Fig. \ref{fig:5.1}, there is a visual discrepancy between the original and protected images of different methods. Existing methods inject distinct perturbations to replace the original samples, whereas ours does not make any modifications since PATFinger relies on the inherent distribution of the dataset rather than artificial trigger patterns. We further provide quantitative results to measure the image quality and the retrieval performance of models in Tab. \ref{tab:2}, which show the difference in retrieval performance after fine-tuning the model on the protected dataset. The results show the superiority of our scheme in exploiting the properties of fingerprints without modifying the original dataset and no additional loss in model retrieval performance. Specifically, we randomly select image-text pairs from clean datasets as inputs in the retrieval performance test. From these average R@1 and R@5, we observe a relatively prominent degradation in image-to-text retrieval performance, \textit{i.e.,} R@5, larger than 5\%. This may be related to the impact of WaNet, DVBW, and UBW on the semantics of pictures. In contrast, the gap between the different methods in text-to-image retrieval (\textit{i.e.,} R@5) is very slight, less than 3\%.

\begin{figure}
\centering
\includegraphics[width=0.9\columnwidth]{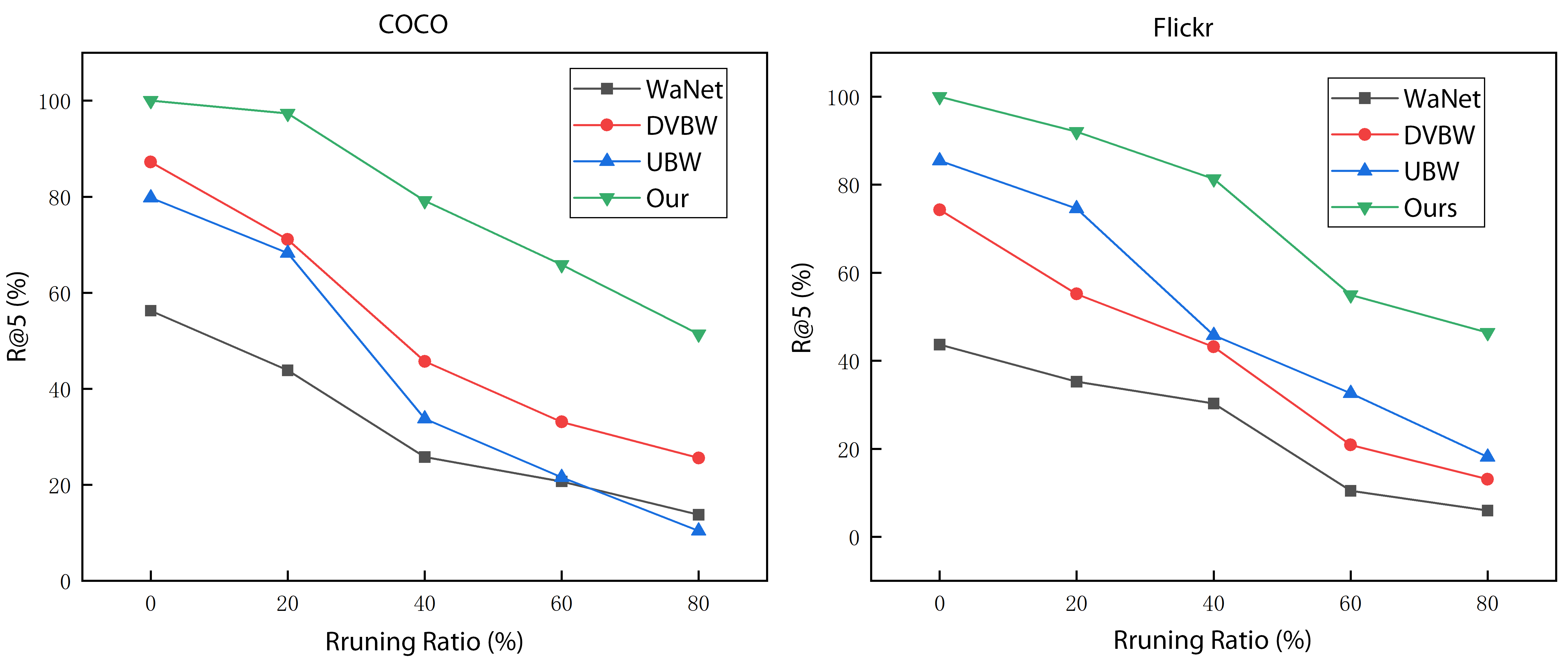}
\setlength{\abovecaptionskip}{0.1cm}
\caption{Robustness of PATFinger and baseline methods against dataset pruning for IT retrieval task.}
\label{fig:5.4}
\end{figure}

\begin{table}[ht]
\centering
\setlength{\abovecaptionskip}{0.1cm}
\small
\caption{Results of ownership verification methods against no defense, fine-tuning, and Clean-CLIP for IT retrieval task. Bold indicates the best results.}
\label{tab:3.2}
\begin{tabular}{ccccccc}
\toprule
\multirow{2}{*}{Dataset} & \multicolumn{2}{c}{No Defense} & \multicolumn{2}{c}{Fine-Tuning} & \multicolumn{2}{c}{Clean-CLIP} \\ 
                         & R@1            & R@5           & R@1            & R@5            & R@1            & R@5           \\ \midrule
WaNet                    &  16.30        & 40.33        &  13.96        & 32.03          & 1.56           & 10.44              \\
DVBW                     & 19.23          & 76.56         & 13.08          & 33.98          & 2.83           & 14.55              \\
UBW                      & 87.79          & 73.33         & 33.00          & 15.72          & 24.70          & 8.59             \\
Ours                      & \textbf{96.09}          & \textbf{100.00}           &\textbf{92.38}           & \textbf{95.60 }         & \textbf{85.35}          & \textbf{88.76 }          \\ 
\bottomrule
\end{tabular}
\end{table}

\begin{table}[ht]
\centering
\setlength{\abovecaptionskip}{0.1cm}
\small
\caption{PATFinger is obtained on surrogate models and evaluated on the other models to evaluate its cross-model transferability for IT retrieval task.}
\label{tab:4}
\begin{tabular}{cccccc}
\toprule
\multirow{2}{*}{Surrogate} & \multirow{2}{*}{Suspicious}  & \multicolumn{2}{c}{COCO}  & \multicolumn{2}{c}{Flickr} \\
 & & R@1         & R@5       & R@1 & R@5 \\ \midrule
\multirow{4}{*}{CLIP\textsubscript{RN101}}    & CLIP\textsubscript{VT16}   &35.44  &60.25 &32.12  &55.85    \\
                                              & CLIP\textsubscript{VT32}   &39.55  &64.25 &39.74  &54.39    \\
                                              & CLIP\textsubscript{VL14}   &32.71  &51.26 &27.92  &51.85    \\
                                              & BLIP                       &28.02  &50.68 &22.16  &52.24    \\ \midrule
\multirow{4}{*}{CLIP\textsubscript{VT16}}     & CLIP\textsubscript{RN101}  &26.75  &56.34 &29.68  &64.74    \\
                                              & CLIP\textsubscript{VT32}   &65.13  &81.15 &62.30  &80.95    \\ 
                                              & CLIP\textsubscript{VL14}   &38.57  &67.87 &38.28  &66.69    \\
                                              & BLIP                       &27.24  &55.46 &32.42  &73.14     \\ \midrule
\multirow{4}{*}{CLIP\textsubscript{VT32}}     & CLIP\textsubscript{RN101}  &60.93  &76.56 &61.81  &79.88     \\
                                              & CLIP\textsubscript{VT16}   &66.60  &81.93 &64.55  &83.59     \\
                                              & CLIP\textsubscript{VL14}   &57.03  &77.34 &56.93  &74.60     \\
                                              & BLIP                       &55.07  &75.39 &59.37  &76.56     \\ \midrule
\multirow{4}{*}{CLIP\textsubscript{VL14}}     & CLIP\textsubscript{RN101}  &39.35  &69.92 &36.71  &57.32     \\
                                              & CLIP\textsubscript{VT16}   &37.20  &63.86 &41.79  &68.45     \\
                                              & CLIP\textsubscript{VT32}   &36.52  &51.36 &33.10  &56.15     \\
                                              & BLIP                       &21.28  &53.80 &30.85  &61.13     \\ \bottomrule
\end{tabular}
\end{table}

\subsection{Analyzing the Robustness under Demanding Conditions (RQ3)}
Based on the violator’s goals, we evaluate the robustness of PATFinger against demanding conditions in the image-text retrieval task. Considering realistic scenarios, adversaries aim to bypass ownership verification by intuitively modifying the dataset through dataset pruning and fine-tuning. 

We first evaluate our PATFinger on scenarios where the adversaries have unrestricted access to selective deletion and reduction of protected datasets, such as, a 20\% pruning ratio means that 20\% of the dataset has been randomly removed. From Fig. \ref{fig:5.4}, we can identify that the R@5 of all methods decreases with the degree of dataset pruning. Notably, our method achieves 45\% R@5 even when the crop ratio reaches 80\%, while the baseline methods have lost their utility.

We further investigate the effectiveness of PATFinger against fine-tuning, involving fine-tuning on a hybrid dataset and the SoTA trigger mitigation method Clean-CLIP~\cite{bansal2023cleanclip}. The protected dataset is COCO, while Flickr, Open-images, and TextCaps are mixed as the hybrid dataset. We evaluate the robustness of all methods under demanding conditions. From Tab. \ref{tab:3.2}, we observe that all ownership verification methods can achieve high retrieval rates in image-text retrieval tasks. Moreover, all baseline methods drop significantly under demanding conditions, especially when the Clean-CLIP defense is used (\textit{e.g.,} R@5 more than 50\% drop). However, in the same case, our PATFinger is robust enough to keep R@5 constant, which demonstrates the superiority of prompt-adapted transferable fingerprinting.

\subsection{Evaluation of Black-Box Transferability (RQ4)}
In this part, we investigate the effectiveness of PATFinger on the black-box setting, which means that the data owner has minimal knowledge about the suspicious model. In this process, we transform the continuous prompts obtained from the surrogate model into discrete textual representations, which serve as input queries for probing the black-box suspicious models. As shown in Tab. \ref{tab:4}, our PATFinger exhibits excellent transferability on COCO and Flickr, with an average  R@5 of over 50\%. Specifically, the PATFinger shows exceptional transferability when both surrogate and suspicious models share similar architectures. For example, our PATFinger can be efficiently transferred between CLIP\textsubscript{VT16} and CLIP\textsubscript{VT32}, possibly due to the proximity of the semantic space. Notably, the CLIP\textsubscript{VT32} outperforms other surrogate models achieving an average R@5 increase of 12\% and being the only model to exceed 75\% utility. Furthermore, despite building upon a multimodal encoder architecture, the PATFinger based on CLIP\textsubscript{VT32} can still effectively adapt to the multimodal mixture of the encoder-decoder model, namely BLIP. Overall, the PATFinger overcomes the drawbacks of continuous prompts by textual constraint, which demonstrates the transferability of PATFinger across black-box VLM.

\subsection{Ablation Study (RQ5)}
In this section, we further investigate the effect of different modules and key hyper-parameters on our proposed PATFinger for IT retrieval task.

\begin{figure}[t]
\centering
\includegraphics[width=0.9\columnwidth]{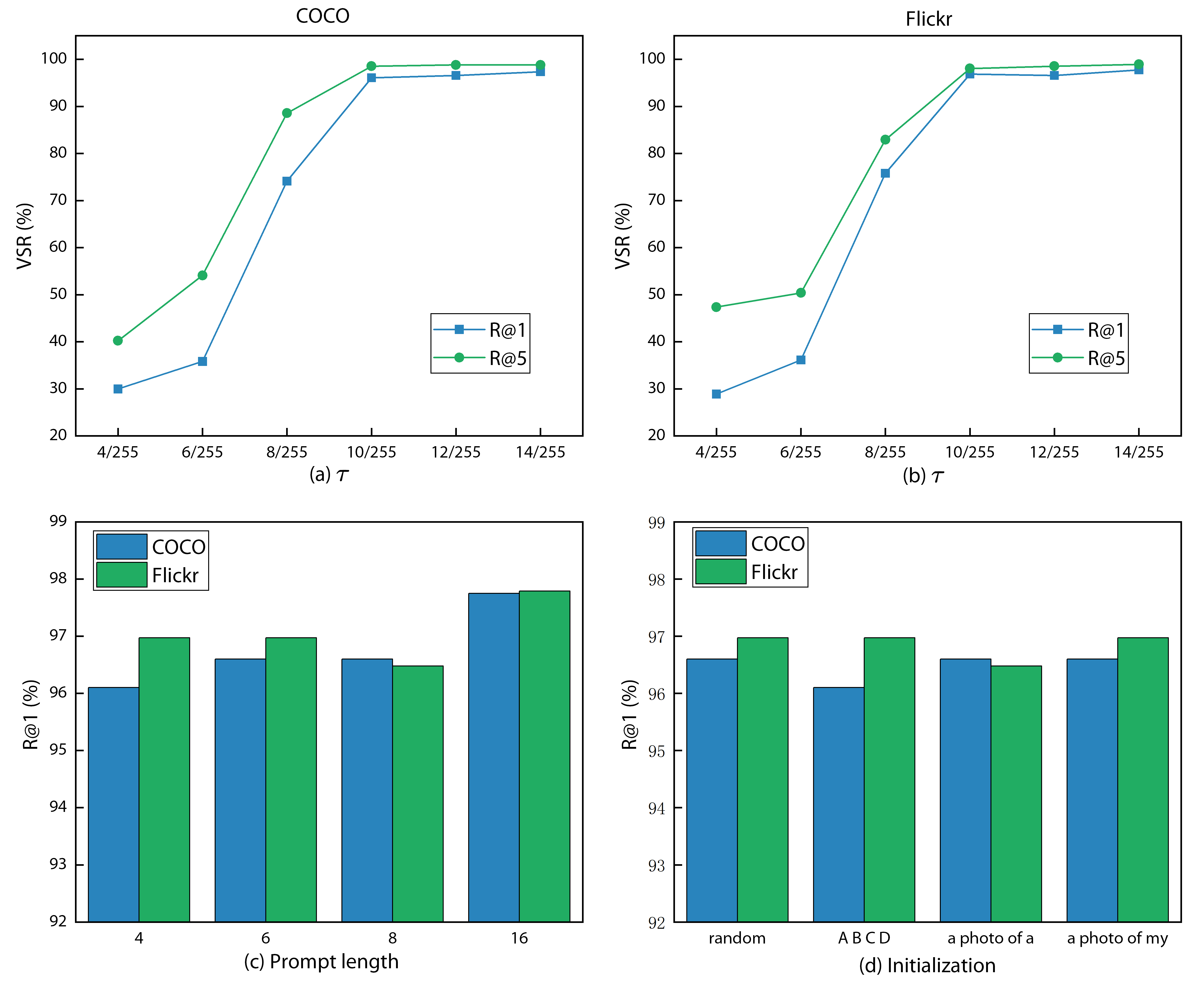}
\caption{The performance of PATFinger under different settings for IT retrieval task. (a) and (b) investigate the effect of perturbation budget. (c) and (d) evaluate the impact of prompt setting.}
\label{fig:5.6}
\end{figure}

\begin{table}[]
\centering
\small
\setlength{\abovecaptionskip}{0.1cm}
\caption{The verification performance under different module combinations for IT retrieval task. The D\textsubscript{intra}, D\textsubscript{inter}, and TC represent the intra-modal discriminator, inter-modal discriminator, and textual constraint, respectively. Bold indicates the best results.}
\label{tab:5.1}
\begin{tabular}{ccc|cc|cc}
\bottomrule
\multicolumn{3}{c|}{Module} & \multicolumn{2}{c|}{CLIP\textsubscript{VT32}} & \multicolumn{2}{c}{CLIP\textsubscript{RN101}} \\ \cline{1-7} 
D\textsubscript{intra}   & D\textsubscript{inter}  & TC  & R@1    & R@5   & R@1    & R@5    \\ \cline{1-7}
$\checkmark$  &$\checkmark$  &$\checkmark$               & \textbf{96.09} & \textbf{100.00}  & \textbf{57.71}  & \textbf{75.29}  \\
$\checkmark$  &              &$\checkmark$               & 74.80  & 87.20  & 37.30  & 54.78  \\
              &$\checkmark$  &$\checkmark$               & 74.12  & 83.69  & 38.28  & 51.66  \\
$\checkmark$  &$\checkmark$  &                           & 95.41  & 100.00 & 27.05  & 36.91  \\
$\checkmark$  &              &                           & 77.73  & 90.52  & 29.78  & 38.67  \\
              &$\checkmark$  &                           & 71.58  & 81.73  & 27.34  & 37.50 \\
\toprule
\end{tabular} 
\end{table}

\noindent\textbf{The effect of each module.}
The results of different module combinations are shown in Tab. \ref{tab:5.1}, which shows that the three modules play different roles. Our results demonstrate that relying on a single discriminator is insufficient for effectively extracting features from multimodal samples, which leads to a diminished verification capability. Specifically, in the white-box setting (\textit{i.e.,} CLIP\textsubscript{VT32}), the double discriminator (\textit{i.e.,} D\textsubscript{intra} and D\textsubscript{inter}) strategy performs best, achieving on average 21.63\% higher than the single discriminator scheme. Interestingly, the textual constraint (TC) module shows no measurable effect in the white-box setting, which may be a result of its limited impact on feature extraction. Moreover, in the black-box setting (\textit{i.e.,} CLIP\textsubscript{RN101}), the performance drops significantly in the absence of the TC module, which indicates that the TC module is crucial for enhanced cross-model transferability.

\noindent\textbf{The effect of the perturbation budget.}
Fig. \ref{fig:5.6}(a) and (b) show the results of evaluating the effect of different perturbation budgets, including 4/255, 6/255, 8/255, 10/255, 12/255, and 14/255. 
We observe that the verification performance increases rapidly and stabilizes when the perturbation reaches 10/255. Specifically, there is a significant improvement by increasing the perturbation budget, but performance decreases once the budget exceeds a certain threshold.
Overall, our PATFinger can achieve a better trade-off between VSR and image visual quality when \(\tau\) is 10/255.

\noindent\textbf{The effect of prompts setting.}
We analyze the effect of the different prompt lengths and initialization strategies. 
Although increasing the length has a minor effect, we can see that a length 4 prompt is sufficient to pull the textual fingerprints closer to the GOP, as shown in Fig. \ref{fig:5.6}(c).
As shown in Fig. \ref{fig:5.6}(d), at the beginning of prompt learning, prompts are initialized to a random string, ``A B C D'', ``a photo of a'' and ``a photo of my'', respectively. Our results show that initialization has a negligible impact on performance.

\section{Conclusion}
In this paper, we reveal that the decision boundaries
of models fine-tuned on the same dataset share consistent perturbation affinities. Based on this understanding, we propose a novel ownership verification framework from the training-free perspective that utilizes inherent model-agnostic distribution attributes (i.e., dataset identifiers) as fingerprints. Our PATFinger comprises the global optimal perturbation and the adaptive prompt to capture the image-text matched decision boundary characteristics in the carefully crafted surrogate model. Our empirical experiments demonstrate that PATFinger outperforms current state-of-the-art methods against the usage of unauthorized multimodal datasets.

\section*{Acknowledgments}
This work was supported by the National Natural Science Foundation of China (62402106, 62372334), the Natural Science Foundation of Jiangsu Province of China (BK20241272), the Fundamental Research Funds for the Central Universities (2242024k30059), and the Start-Up Research Fund of Southeast University (RF1028623129).

\bibliographystyle{ACM-Reference-Format}
\bibliography{sample-base}


\begin{thebibliography}{62}


\ifx \showCODEN    \undefined \def \showCODEN     #1{\unskip}     \fi
\ifx \showISBNx    \undefined \def \showISBNx     #1{\unskip}     \fi
\ifx \showISBNxiii \undefined \def \showISBNxiii  #1{\unskip}     \fi
\ifx \showISSN     \undefined \def \showISSN      #1{\unskip}     \fi
\ifx \showLCCN     \undefined \def \showLCCN      #1{\unskip}     \fi
\ifx \shownote     \undefined \def \shownote      #1{#1}          \fi
\ifx \showarticletitle \undefined \def \showarticletitle #1{#1}   \fi
\ifx \showURL      \undefined \def \showURL       {\relax}        \fi
\providecommand\bibfield[2]{#2}
\providecommand\bibinfo[2]{#2}
\providecommand\natexlab[1]{#1}
\providecommand\showeprint[2][]{arXiv:#2}

\bibitem[Bansal et~al\mbox{.}(2023)]%
        {bansal2023cleanclip}
\bibfield{author}{\bibinfo{person}{Hritik Bansal}, \bibinfo{person}{Nishad Singhi}, \bibinfo{person}{Yu Yang}, \bibinfo{person}{Fan Yin}, \bibinfo{person}{Aditya Grover}, {and} \bibinfo{person}{Kai-Wei Chang}.} \bibinfo{year}{2023}\natexlab{}.
\newblock \showarticletitle{Cleanclip: Mitigating data poisoning attacks in multimodal contrastive learning}. In \bibinfo{booktitle}{\emph{Proceedings of the IEEE/CVF International Conference on Computer Vision}}. \bibinfo{pages}{112--123}.
\newblock


\bibitem[Cao et~al\mbox{.}(2021)]%
        {cao2021ipguard}
\bibfield{author}{\bibinfo{person}{Xiaoyu Cao}, \bibinfo{person}{Jinyuan Jia}, {and} \bibinfo{person}{Neil~Zhenqiang Gong}.} \bibinfo{year}{2021}\natexlab{}.
\newblock \showarticletitle{IPGuard: Protecting intellectual property of deep neural networks via fingerprinting the classification boundary}. In \bibinfo{booktitle}{\emph{Proceedings of the 2021 ACM asia conference on computer and communications security}}. \bibinfo{pages}{14--25}.
\newblock


\bibitem[Chen et~al\mbox{.}(2023b)]%
        {chen2023vlp}
\bibfield{author}{\bibinfo{person}{Fei-Long Chen}, \bibinfo{person}{Du-Zhen Zhang}, \bibinfo{person}{Ming-Lun Han}, \bibinfo{person}{Xiu-Yi Chen}, \bibinfo{person}{Jing Shi}, \bibinfo{person}{Shuang Xu}, {and} \bibinfo{person}{Bo Xu}.} \bibinfo{year}{2023}\natexlab{b}.
\newblock \showarticletitle{Vlp: A survey on vision-language pre-training}.
\newblock \bibinfo{journal}{\emph{Machine Intelligence Research}} \bibinfo{volume}{20}, \bibinfo{number}{1} (\bibinfo{year}{2023}), \bibinfo{pages}{38--56}.
\newblock


\bibitem[Chen et~al\mbox{.}(2023a)]%
        {chen2023rethinking}
\bibfield{author}{\bibinfo{person}{Weijing Chen}, \bibinfo{person}{Linli Yao}, {and} \bibinfo{person}{Qin Jin}.} \bibinfo{year}{2023}\natexlab{a}.
\newblock \showarticletitle{Rethinking benchmarks for cross-modal image-text retrieval}. In \bibinfo{booktitle}{\emph{Proceedings of the 46th International ACM SIGIR Conference on Research and Development in Information Retrieval}}. \bibinfo{pages}{1241--1251}.
\newblock


\bibitem[Dziedzic et~al\mbox{.}(2022)]%
        {dziedzic2022dataset}
\bibfield{author}{\bibinfo{person}{Adam Dziedzic}, \bibinfo{person}{Haonan Duan}, \bibinfo{person}{Muhammad~Ahmad Kaleem}, \bibinfo{person}{Nikita Dhawan}, \bibinfo{person}{Jonas Guan}, \bibinfo{person}{Yannis Cattan}, \bibinfo{person}{Franziska Boenisch}, {and} \bibinfo{person}{Nicolas Papernot}.} \bibinfo{year}{2022}\natexlab{}.
\newblock \showarticletitle{Dataset inference for self-supervised models}.
\newblock \bibinfo{journal}{\emph{Advances in Neural Information Processing Systems}}  \bibinfo{volume}{35} (\bibinfo{year}{2022}), \bibinfo{pages}{12058--12070}.
\newblock


\bibitem[Fu et~al\mbox{.}(2023)]%
        {fu2023effectiveness}
\bibfield{author}{\bibinfo{person}{Zihao Fu}, \bibinfo{person}{Haoran Yang}, \bibinfo{person}{Anthony Man-Cho So}, \bibinfo{person}{Wai Lam}, \bibinfo{person}{Lidong Bing}, {and} \bibinfo{person}{Nigel Collier}.} \bibinfo{year}{2023}\natexlab{}.
\newblock \showarticletitle{On the effectiveness of parameter-efficient fine-tuning}. In \bibinfo{booktitle}{\emph{Proceedings of the AAAI conference on artificial intelligence}}, Vol.~\bibinfo{volume}{37}. \bibinfo{pages}{12799--12807}.
\newblock


\bibitem[Gan et~al\mbox{.}(2020)]%
        {gan2020large}
\bibfield{author}{\bibinfo{person}{Zhe Gan}, \bibinfo{person}{Yen-Chun Chen}, \bibinfo{person}{Linjie Li}, \bibinfo{person}{Chen Zhu}, \bibinfo{person}{Yu Cheng}, {and} \bibinfo{person}{Jingjing Liu}.} \bibinfo{year}{2020}\natexlab{}.
\newblock \showarticletitle{Large-scale adversarial training for vision-and-language representation learning}.
\newblock \bibinfo{journal}{\emph{Advances in Neural Information Processing Systems}}  \bibinfo{volume}{33} (\bibinfo{year}{2020}), \bibinfo{pages}{6616--6628}.
\newblock


\bibitem[Guo et~al\mbox{.}(2024a)]%
        {guo2024zero}
\bibfield{author}{\bibinfo{person}{Junfeng Guo}, \bibinfo{person}{Yiming Li}, \bibinfo{person}{Ruibo Chen}, \bibinfo{person}{Yihan Wu}, \bibinfo{person}{Chenxi Liu}, {and} \bibinfo{person}{Heng Huang}.} \bibinfo{year}{2024}\natexlab{a}.
\newblock \showarticletitle{ZeroMark: Towards Dataset Ownership Verification without Disclosing Watermarks}. In \bibinfo{booktitle}{\emph{NeurIPS}}.
\newblock


\bibitem[Guo et~al\mbox{.}(2024b)]%
        {guo2024domain}
\bibfield{author}{\bibinfo{person}{Junfeng Guo}, \bibinfo{person}{Yiming Li}, \bibinfo{person}{Lixu Wang}, \bibinfo{person}{Shu-Tao Xia}, \bibinfo{person}{Heng Huang}, \bibinfo{person}{Cong Liu}, {and} \bibinfo{person}{Bo Li}.} \bibinfo{year}{2024}\natexlab{b}.
\newblock \showarticletitle{Domain watermark: Effective and harmless dataset copyright protection is closed at hand}.
\newblock \bibinfo{journal}{\emph{Advances in Neural Information Processing Systems}}  \bibinfo{volume}{36} (\bibinfo{year}{2024}).
\newblock


\bibitem[He et~al\mbox{.}(2016)]%
        {he2016deep}
\bibfield{author}{\bibinfo{person}{Kaiming He}, \bibinfo{person}{Xiangyu Zhang}, \bibinfo{person}{Shaoqing Ren}, {and} \bibinfo{person}{Jian Sun}.} \bibinfo{year}{2016}\natexlab{}.
\newblock \showarticletitle{Deep residual learning for image recognition}. In \bibinfo{booktitle}{\emph{Proceedings of the IEEE conference on computer vision and pattern recognition}}. \bibinfo{pages}{770--778}.
\newblock


\bibitem[Hong et~al\mbox{.}(2021)]%
        {hong2021gilbert}
\bibfield{author}{\bibinfo{person}{Weixiang Hong}, \bibinfo{person}{Kaixiang Ji}, \bibinfo{person}{Jiajia Liu}, \bibinfo{person}{Jian Wang}, \bibinfo{person}{Jingdong Chen}, {and} \bibinfo{person}{Wei Chu}.} \bibinfo{year}{2021}\natexlab{}.
\newblock \showarticletitle{Gilbert: Generative vision-language pre-training for image-text retrieval}. In \bibinfo{booktitle}{\emph{Proceedings of the 44th International ACM SIGIR Conference on Research and Development in Information Retrieval}}. \bibinfo{pages}{1379--1388}.
\newblock


\bibitem[Hore and Ziou(2010)]%
        {hore2010image}
\bibfield{author}{\bibinfo{person}{Alain Hore} {and} \bibinfo{person}{Djemel Ziou}.} \bibinfo{year}{2010}\natexlab{}.
\newblock \showarticletitle{Image quality metrics: PSNR vs. SSIM}. In \bibinfo{booktitle}{\emph{2010 20th international conference on pattern recognition}}. IEEE, \bibinfo{pages}{2366--2369}.
\newblock


\bibitem[Hu et~al\mbox{.}(2022)]%
        {Hu_2022_CVPR}
\bibfield{author}{\bibinfo{person}{Xiaowei Hu}, \bibinfo{person}{Zhe Gan}, \bibinfo{person}{Jianfeng Wang}, \bibinfo{person}{Zhengyuan Yang}, \bibinfo{person}{Zicheng Liu}, \bibinfo{person}{Yumao Lu}, {and} \bibinfo{person}{Lijuan Wang}.} \bibinfo{year}{2022}\natexlab{}.
\newblock \showarticletitle{Scaling Up Vision-Language Pre-Training for Image Captioning}. In \bibinfo{booktitle}{\emph{Proceedings of the IEEE/CVF Conference on Computer Vision and Pattern Recognition (CVPR)}}. \bibinfo{pages}{17980--17989}.
\newblock


\bibitem[Huang et~al\mbox{.}(2024)]%
        {huang2024texture}
\bibfield{author}{\bibinfo{person}{Yihao Huang}, \bibinfo{person}{Qing Guo}, \bibinfo{person}{Felix Juefei-Xu}, \bibinfo{person}{Ming Hu}, \bibinfo{person}{Xiaojun Jia}, \bibinfo{person}{Xiaochun Cao}, \bibinfo{person}{Geguang Pu}, {and} \bibinfo{person}{Yang Liu}.} \bibinfo{year}{2024}\natexlab{}.
\newblock \showarticletitle{Texture re-scalable universal adversarial perturbation}.
\newblock \bibinfo{journal}{\emph{IEEE Transactions on Information Forensics and Security}} (\bibinfo{year}{2024}).
\newblock


\bibitem[Jia et~al\mbox{.}(2021)]%
        {jia2021scaling}
\bibfield{author}{\bibinfo{person}{Chao Jia}, \bibinfo{person}{Yinfei Yang}, \bibinfo{person}{Ye Xia}, \bibinfo{person}{Yi-Ting Chen}, \bibinfo{person}{Zarana Parekh}, \bibinfo{person}{Hieu Pham}, \bibinfo{person}{Quoc Le}, \bibinfo{person}{Yun-Hsuan Sung}, \bibinfo{person}{Zhen Li}, {and} \bibinfo{person}{Tom Duerig}.} \bibinfo{year}{2021}\natexlab{}.
\newblock \showarticletitle{Scaling up visual and vision-language representation learning with noisy text supervision}. In \bibinfo{booktitle}{\emph{International conference on machine learning}}. PMLR, \bibinfo{pages}{4904--4916}.
\newblock


\bibitem[Jia et~al\mbox{.}(2022a)]%
        {jia2022partial}
\bibfield{author}{\bibinfo{person}{Ju Jia}, \bibinfo{person}{Meng Luo}, \bibinfo{person}{Siqi Ma}, {and} \bibinfo{person}{Lina Wang}.} \bibinfo{year}{2022}\natexlab{a}.
\newblock \showarticletitle{Partial knowledge transfer in visual recognition systems via joint loss-aware consistency learning}.
\newblock \bibinfo{journal}{\emph{IEEE Transactions on Industrial Informatics}} \bibinfo{volume}{18}, \bibinfo{number}{11} (\bibinfo{year}{2022}), \bibinfo{pages}{7463--7474}.
\newblock


\bibitem[Jia et~al\mbox{.}(2022b)]%
        {jia2022consensus}
\bibfield{author}{\bibinfo{person}{Ju Jia}, \bibinfo{person}{Meng Luo}, \bibinfo{person}{Siqi Ma}, \bibinfo{person}{Lina Wang}, {and} \bibinfo{person}{Yang Liu}.} \bibinfo{year}{2022}\natexlab{b}.
\newblock \showarticletitle{Consensus-clustering-based automatic distribution matching for cross-domain image steganalysis}.
\newblock \bibinfo{journal}{\emph{IEEE Transactions on Knowledge and Data Engineering}} \bibinfo{volume}{35}, \bibinfo{number}{6} (\bibinfo{year}{2022}), \bibinfo{pages}{5665--5679}.
\newblock


\bibitem[Jia et~al\mbox{.}(2022c)]%
        {jia2022subnetwork}
\bibfield{author}{\bibinfo{person}{Ju Jia}, \bibinfo{person}{Yueming Wu}, \bibinfo{person}{Anran Li}, \bibinfo{person}{Siqi Ma}, {and} \bibinfo{person}{Yang Liu}.} \bibinfo{year}{2022}\natexlab{c}.
\newblock \showarticletitle{Subnetwork-lossless robust watermarking for hostile theft attacks in deep transfer learning models}.
\newblock \bibinfo{journal}{\emph{IEEE transactions on dependable and secure computing}} (\bibinfo{year}{2022}).
\newblock


\bibitem[Jia et~al\mbox{.}(2024)]%
        {jia2024semantic}
\bibfield{author}{\bibinfo{person}{Xiaojun Jia}, \bibinfo{person}{Sensen Gao}, \bibinfo{person}{Qing Guo}, \bibinfo{person}{Ke Ma}, \bibinfo{person}{Yihao Huang}, \bibinfo{person}{Simeng Qin}, \bibinfo{person}{Yang Liu}, {and} \bibinfo{person}{Xiaochun Cao}.} \bibinfo{year}{2024}\natexlab{}.
\newblock \showarticletitle{Semantic-Aligned Adversarial Evolution Triangle for High-Transferability Vision-Language Attack}.
\newblock \bibinfo{journal}{\emph{arXiv preprint arXiv:2411.02669}} (\bibinfo{year}{2024}).
\newblock


\bibitem[Jia et~al\mbox{.}(2020)]%
        {jia2020adv}
\bibfield{author}{\bibinfo{person}{Xiaojun Jia}, \bibinfo{person}{Xingxing Wei}, \bibinfo{person}{Xiaochun Cao}, {and} \bibinfo{person}{Xiaoguang Han}.} \bibinfo{year}{2020}\natexlab{}.
\newblock \showarticletitle{Adv-watermark: A novel watermark perturbation for adversarial examples}. In \bibinfo{booktitle}{\emph{Proceedings of the 28th ACM international conference on multimedia}}. \bibinfo{pages}{1579--1587}.
\newblock


\bibitem[Khashabi et~al\mbox{.}(2021)]%
        {khashabi2021prompt}
\bibfield{author}{\bibinfo{person}{Daniel Khashabi}, \bibinfo{person}{Shane Lyu}, \bibinfo{person}{Sewon Min}, \bibinfo{person}{Lianhui Qin}, \bibinfo{person}{Kyle Richardson}, \bibinfo{person}{Sean Welleck}, \bibinfo{person}{Hannaneh Hajishirzi}, \bibinfo{person}{Tushar Khot}, \bibinfo{person}{Ashish Sabharwal}, \bibinfo{person}{Sameer Singh}, {et~al\mbox{.}}} \bibinfo{year}{2021}\natexlab{}.
\newblock \showarticletitle{Prompt waywardness: The curious case of discretized interpretation of continuous prompts}.
\newblock \bibinfo{journal}{\emph{arXiv preprint arXiv:2112.08348}} (\bibinfo{year}{2021}).
\newblock


\bibitem[Kingma(2014)]%
        {kingma2014adam}
\bibfield{author}{\bibinfo{person}{Diederik~P Kingma}.} \bibinfo{year}{2014}\natexlab{}.
\newblock \showarticletitle{Adam: A method for stochastic optimization}.
\newblock \bibinfo{journal}{\emph{arXiv preprint arXiv:1412.6980}} (\bibinfo{year}{2014}).
\newblock


\bibitem[Kuznetsova et~al\mbox{.}(2020)]%
        {kuznetsova2020open}
\bibfield{author}{\bibinfo{person}{Alina Kuznetsova}, \bibinfo{person}{Hassan Rom}, \bibinfo{person}{Neil Alldrin}, \bibinfo{person}{Jasper Uijlings}, \bibinfo{person}{Ivan Krasin}, \bibinfo{person}{Jordi Pont-Tuset}, \bibinfo{person}{Shahab Kamali}, \bibinfo{person}{Stefan Popov}, \bibinfo{person}{Matteo Malloci}, \bibinfo{person}{Alexander Kolesnikov}, {et~al\mbox{.}}} \bibinfo{year}{2020}\natexlab{}.
\newblock \showarticletitle{The open images dataset v4: Unified image classification, object detection, and visual relationship detection at scale}.
\newblock \bibinfo{journal}{\emph{International journal of computer vision}} \bibinfo{volume}{128}, \bibinfo{number}{7} (\bibinfo{year}{2020}), \bibinfo{pages}{1956--1981}.
\newblock


\bibitem[Lauren{\c{c}}on et~al\mbox{.}(2024)]%
        {laurenccon2024matters}
\bibfield{author}{\bibinfo{person}{Hugo Lauren{\c{c}}on}, \bibinfo{person}{L{\'e}o Tronchon}, \bibinfo{person}{Matthieu Cord}, {and} \bibinfo{person}{Victor Sanh}.} \bibinfo{year}{2024}\natexlab{}.
\newblock \showarticletitle{What matters when building vision-language models?}
\newblock \bibinfo{journal}{\emph{arXiv preprint arXiv:2405.02246}} (\bibinfo{year}{2024}).
\newblock


\bibitem[Li et~al\mbox{.}(2024a)]%
        {li2024towards}
\bibfield{author}{\bibinfo{person}{Boheng Li}, \bibinfo{person}{Yanhao Wei}, \bibinfo{person}{Yankai Fu}, \bibinfo{person}{Zhenting Wang}, \bibinfo{person}{Yiming Li}, \bibinfo{person}{Jie Zhang}, \bibinfo{person}{Run Wang}, {and} \bibinfo{person}{Tianwei Zhang}.} \bibinfo{year}{2024}\natexlab{a}.
\newblock \showarticletitle{Towards Reliable Verification of Unauthorized Data Usage in Personalized Text-to-Image Diffusion Models}.
\newblock \bibinfo{journal}{\emph{arXiv preprint arXiv:2410.10437}} (\bibinfo{year}{2024}).
\newblock


\bibitem[Li et~al\mbox{.}(2024b)]%
        {li2024llava}
\bibfield{author}{\bibinfo{person}{Chunyuan Li}, \bibinfo{person}{Cliff Wong}, \bibinfo{person}{Sheng Zhang}, \bibinfo{person}{Naoto Usuyama}, \bibinfo{person}{Haotian Liu}, \bibinfo{person}{Jianwei Yang}, \bibinfo{person}{Tristan Naumann}, \bibinfo{person}{Hoifung Poon}, {and} \bibinfo{person}{Jianfeng Gao}.} \bibinfo{year}{2024}\natexlab{b}.
\newblock \showarticletitle{Llava-med: Training a large language-and-vision assistant for biomedicine in one day}.
\newblock \bibinfo{journal}{\emph{Advances in Neural Information Processing Systems}}  \bibinfo{volume}{36} (\bibinfo{year}{2024}).
\newblock


\bibitem[Li et~al\mbox{.}(2022b)]%
        {li2022blip}
\bibfield{author}{\bibinfo{person}{Junnan Li}, \bibinfo{person}{Dongxu Li}, \bibinfo{person}{Caiming Xiong}, {and} \bibinfo{person}{Steven Hoi}.} \bibinfo{year}{2022}\natexlab{b}.
\newblock \showarticletitle{Blip: Bootstrapping language-image pre-training for unified vision-language understanding and generation}. In \bibinfo{booktitle}{\emph{International conference on machine learning}}. PMLR, \bibinfo{pages}{12888--12900}.
\newblock


\bibitem[Li et~al\mbox{.}(2022a)]%
        {li2022untargeted}
\bibfield{author}{\bibinfo{person}{Yiming Li}, \bibinfo{person}{Yang Bai}, \bibinfo{person}{Yong Jiang}, \bibinfo{person}{Yong Yang}, \bibinfo{person}{Shu-Tao Xia}, {and} \bibinfo{person}{Bo Li}.} \bibinfo{year}{2022}\natexlab{a}.
\newblock \showarticletitle{Untargeted Backdoor Watermark: Towards Harmless and Stealthy Dataset Copyright Protection}. In \bibinfo{booktitle}{\emph{NeurIPS}}.
\newblock


\bibitem[Li et~al\mbox{.}(2023)]%
        {li2023black}
\bibfield{author}{\bibinfo{person}{Yiming Li}, \bibinfo{person}{Mingyan Zhu}, \bibinfo{person}{Xue Yang}, \bibinfo{person}{Yong Jiang}, \bibinfo{person}{Tao Wei}, {and} \bibinfo{person}{Shu-Tao Xia}.} \bibinfo{year}{2023}\natexlab{}.
\newblock \showarticletitle{Black-box Dataset Ownership Verification via Backdoor Watermarking}.
\newblock \bibinfo{journal}{\emph{IEEE Transactions on Information Forensics and Security}} (\bibinfo{year}{2023}).
\newblock


\bibitem[Lin et~al\mbox{.}(2014)]%
        {lin2014microsoft}
\bibfield{author}{\bibinfo{person}{Tsung-Yi Lin}, \bibinfo{person}{Michael Maire}, \bibinfo{person}{Serge Belongie}, \bibinfo{person}{James Hays}, \bibinfo{person}{Pietro Perona}, \bibinfo{person}{Deva Ramanan}, \bibinfo{person}{Piotr Doll{\'a}r}, {and} \bibinfo{person}{C~Lawrence Zitnick}.} \bibinfo{year}{2014}\natexlab{}.
\newblock \showarticletitle{Microsoft coco: Common objects in context}. In \bibinfo{booktitle}{\emph{Computer Vision--ECCV 2014: 13th European Conference, Zurich, Switzerland, September 6-12, 2014, Proceedings, Part V 13}}. Springer, \bibinfo{pages}{740--755}.
\newblock


\bibitem[Liu et~al\mbox{.}(2022)]%
        {liu2022your}
\bibfield{author}{\bibinfo{person}{Gaoyang Liu}, \bibinfo{person}{Tianlong Xu}, \bibinfo{person}{Xiaoqiang Ma}, {and} \bibinfo{person}{Chen Wang}.} \bibinfo{year}{2022}\natexlab{}.
\newblock \showarticletitle{Your model trains on my data? Protecting intellectual property of training data via membership fingerprint authentication}.
\newblock \bibinfo{journal}{\emph{IEEE Transactions on Information Forensics and Security}}  \bibinfo{volume}{17} (\bibinfo{year}{2022}), \bibinfo{pages}{1024--1037}.
\newblock


\bibitem[Liu et~al\mbox{.}(2023)]%
        {liu2023pre}
\bibfield{author}{\bibinfo{person}{Pengfei Liu}, \bibinfo{person}{Weizhe Yuan}, \bibinfo{person}{Jinlan Fu}, \bibinfo{person}{Zhengbao Jiang}, \bibinfo{person}{Hiroaki Hayashi}, {and} \bibinfo{person}{Graham Neubig}.} \bibinfo{year}{2023}\natexlab{}.
\newblock \showarticletitle{Pre-train, prompt, and predict: A systematic survey of prompting methods in natural language processing}.
\newblock \bibinfo{journal}{\emph{Comput. Surveys}} \bibinfo{volume}{55}, \bibinfo{number}{9} (\bibinfo{year}{2023}), \bibinfo{pages}{1--35}.
\newblock


\bibitem[Liu et~al\mbox{.}(2025)]%
        {liu2025persguard}
\bibfield{author}{\bibinfo{person}{Xinwei Liu}, \bibinfo{person}{Xiaojun Jia}, \bibinfo{person}{Yuan Xun}, \bibinfo{person}{Hua Zhang}, {and} \bibinfo{person}{Xiaochun Cao}.} \bibinfo{year}{2025}\natexlab{}.
\newblock \showarticletitle{PersGuard: Preventing Malicious Personalization via Backdoor Attacks on Pre-trained Text-to-Image Diffusion Models}.
\newblock \bibinfo{journal}{\emph{arXiv preprint arXiv:2502.16167}} (\bibinfo{year}{2025}).
\newblock


\bibitem[Lukas et~al\mbox{.}({[n.\,d.]})]%
        {lukasdeep}
\bibfield{author}{\bibinfo{person}{Nils Lukas}, \bibinfo{person}{Yuxuan Zhang}, {and} \bibinfo{person}{Florian Kerschbaum}.} \bibinfo{year}{[n.\,d.]}\natexlab{}.
\newblock \showarticletitle{Deep Neural Network Fingerprinting by Conferrable Adversarial Examples}. In \bibinfo{booktitle}{\emph{International Conference on Learning Representations}}.
\newblock


\bibitem[Maini et~al\mbox{.}(2021)]%
        {maini2021dataset}
\bibfield{author}{\bibinfo{person}{Pratyush Maini}, \bibinfo{person}{Mohammad Yaghini}, {and} \bibinfo{person}{Nicolas Papernot}.} \bibinfo{year}{2021}\natexlab{}.
\newblock \showarticletitle{Dataset inference: Ownership resolution in machine learning}.
\newblock \bibinfo{journal}{\emph{arXiv preprint arXiv:2104.10706}} (\bibinfo{year}{2021}).
\newblock


\bibitem[McKinzie et~al\mbox{.}(2025)]%
        {mckinzie2025mm1}
\bibfield{author}{\bibinfo{person}{Brandon McKinzie}, \bibinfo{person}{Zhe Gan}, \bibinfo{person}{Jean-Philippe Fauconnier}, \bibinfo{person}{Sam Dodge}, \bibinfo{person}{Bowen Zhang}, \bibinfo{person}{Philipp Dufter}, \bibinfo{person}{Dhruti Shah}, \bibinfo{person}{Xianzhi Du}, \bibinfo{person}{Futang Peng}, \bibinfo{person}{Anton Belyi}, {et~al\mbox{.}}} \bibinfo{year}{2025}\natexlab{}.
\newblock \showarticletitle{MM1: methods, analysis and insights from multimodal LLM pre-training}. In \bibinfo{booktitle}{\emph{European Conference on Computer Vision}}. Springer, \bibinfo{pages}{304--323}.
\newblock


\bibitem[Moosavi-Dezfooli et~al\mbox{.}(2017)]%
        {moosavi2017universal}
\bibfield{author}{\bibinfo{person}{Seyed-Mohsen Moosavi-Dezfooli}, \bibinfo{person}{Alhussein Fawzi}, \bibinfo{person}{Omar Fawzi}, {and} \bibinfo{person}{Pascal Frossard}.} \bibinfo{year}{2017}\natexlab{}.
\newblock \showarticletitle{Universal adversarial perturbations}. In \bibinfo{booktitle}{\emph{Proceedings of the IEEE conference on computer vision and pattern recognition}}. \bibinfo{pages}{1765--1773}.
\newblock


\bibitem[Nguyen and Tran(2021)]%
        {nguyen2021wanet}
\bibfield{author}{\bibinfo{person}{Anh Nguyen} {and} \bibinfo{person}{Anh Tran}.} \bibinfo{year}{2021}\natexlab{}.
\newblock \showarticletitle{Wanet--imperceptible warping-based backdoor attack}.
\newblock \bibinfo{journal}{\emph{arXiv preprint arXiv:2102.10369}} (\bibinfo{year}{2021}).
\newblock


\bibitem[Oord et~al\mbox{.}(2018)]%
        {oord2018representation}
\bibfield{author}{\bibinfo{person}{Aaron van~den Oord}, \bibinfo{person}{Yazhe Li}, {and} \bibinfo{person}{Oriol Vinyals}.} \bibinfo{year}{2018}\natexlab{}.
\newblock \showarticletitle{Representation learning with contrastive predictive coding}.
\newblock \bibinfo{journal}{\emph{arXiv preprint arXiv:1807.03748}} (\bibinfo{year}{2018}).
\newblock


\bibitem[Pei et~al\mbox{.}(2025)]%
        {pei2025selfprompt}
\bibfield{author}{\bibinfo{person}{Aihua Pei}, \bibinfo{person}{Zehua Yang}, \bibinfo{person}{Shunan Zhu}, \bibinfo{person}{Ruoxi Cheng}, {and} \bibinfo{person}{Ju Jia}.} \bibinfo{year}{2025}\natexlab{}.
\newblock \showarticletitle{SelfPrompt: Autonomously Evaluating LLM Robustness via Domain-Constrained Knowledge Guidelines and Refined Adversarial Prompts}. In \bibinfo{booktitle}{\emph{Proceedings of the 31st International Conference on Computational Linguistics}}. \bibinfo{pages}{6840--6854}.
\newblock


\bibitem[Peng et~al\mbox{.}(2022)]%
        {peng2022fingerprinting}
\bibfield{author}{\bibinfo{person}{Zirui Peng}, \bibinfo{person}{Shaofeng Li}, \bibinfo{person}{Guoxing Chen}, \bibinfo{person}{Cheng Zhang}, \bibinfo{person}{Haojin Zhu}, {and} \bibinfo{person}{Minhui Xue}.} \bibinfo{year}{2022}\natexlab{}.
\newblock \showarticletitle{Fingerprinting deep neural networks globally via universal adversarial perturbations}. In \bibinfo{booktitle}{\emph{Proceedings of the IEEE/CVF conference on computer vision and pattern recognition}}. \bibinfo{pages}{13430--13439}.
\newblock


\bibitem[Plummer et~al\mbox{.}(2017)]%
        {flickrentitiesijcv}
\bibfield{author}{\bibinfo{person}{Bryan~A. Plummer}, \bibinfo{person}{Liwei Wang}, \bibinfo{person}{Christopher~M. Cervantes}, \bibinfo{person}{Juan~C. Caicedo}, \bibinfo{person}{Julia Hockenmaier}, {and} \bibinfo{person}{Svetlana Lazebnik}.} \bibinfo{year}{2017}\natexlab{}.
\newblock \showarticletitle{Flickr30K Entities: Collecting Region-to-Phrase Correspondences for Richer Image-to-Sentence Models}.
\newblock \bibinfo{journal}{\emph{IJCV}} \bibinfo{volume}{123}, \bibinfo{number}{1} (\bibinfo{year}{2017}), \bibinfo{pages}{74--93}.
\newblock


\bibitem[Radford et~al\mbox{.}(2021)]%
        {radford2021learning}
\bibfield{author}{\bibinfo{person}{Alec Radford}, \bibinfo{person}{Jong~Wook Kim}, \bibinfo{person}{Chris Hallacy}, \bibinfo{person}{Aditya Ramesh}, \bibinfo{person}{Gabriel Goh}, \bibinfo{person}{Sandhini Agarwal}, \bibinfo{person}{Girish Sastry}, \bibinfo{person}{Amanda Askell}, \bibinfo{person}{Pamela Mishkin}, \bibinfo{person}{Jack Clark}, {et~al\mbox{.}}} \bibinfo{year}{2021}\natexlab{}.
\newblock \showarticletitle{Learning transferable visual models from natural language supervision}. In \bibinfo{booktitle}{\emph{International conference on machine learning}}. PMLR, \bibinfo{pages}{8748--8763}.
\newblock


\bibitem[Schuhmann et~al\mbox{.}(2022)]%
        {schuhmann2022laion}
\bibfield{author}{\bibinfo{person}{Christoph Schuhmann}, \bibinfo{person}{Romain Beaumont}, \bibinfo{person}{Richard Vencu}, \bibinfo{person}{Cade Gordon}, \bibinfo{person}{Ross Wightman}, \bibinfo{person}{Mehdi Cherti}, \bibinfo{person}{Theo Coombes}, \bibinfo{person}{Aarush Katta}, \bibinfo{person}{Clayton Mullis}, \bibinfo{person}{Mitchell Wortsman}, {et~al\mbox{.}}} \bibinfo{year}{2022}\natexlab{}.
\newblock \showarticletitle{Laion-5b: An open large-scale dataset for training next generation image-text models}.
\newblock \bibinfo{journal}{\emph{Advances in Neural Information Processing Systems}}  \bibinfo{volume}{35} (\bibinfo{year}{2022}), \bibinfo{pages}{25278--25294}.
\newblock


\bibitem[Schuhmann et~al\mbox{.}(2021)]%
        {schuhmann2021laion}
\bibfield{author}{\bibinfo{person}{Christoph Schuhmann}, \bibinfo{person}{Richard Vencu}, \bibinfo{person}{Romain Beaumont}, \bibinfo{person}{Robert Kaczmarczyk}, \bibinfo{person}{Clayton Mullis}, \bibinfo{person}{Aarush Katta}, \bibinfo{person}{Theo Coombes}, \bibinfo{person}{Jenia Jitsev}, {and} \bibinfo{person}{Aran Komatsuzaki}.} \bibinfo{year}{2021}\natexlab{}.
\newblock \showarticletitle{Laion-400m: Open dataset of clip-filtered 400 million image-text pairs}.
\newblock \bibinfo{journal}{\emph{arXiv preprint arXiv:2111.02114}} (\bibinfo{year}{2021}).
\newblock


\bibitem[Sidorov et~al\mbox{.}(2020)]%
        {sidorov2020textcaps}
\bibfield{author}{\bibinfo{person}{Oleksii Sidorov}, \bibinfo{person}{Ronghang Hu}, \bibinfo{person}{Marcus Rohrbach}, {and} \bibinfo{person}{Amanpreet Singh}.} \bibinfo{year}{2020}\natexlab{}.
\newblock \showarticletitle{Textcaps: a dataset for image captioning with reading comprehension}. In \bibinfo{booktitle}{\emph{Computer Vision--ECCV 2020: 16th European Conference, Glasgow, UK, August 23--28, 2020, Proceedings, Part II 16}}. Springer, \bibinfo{pages}{742--758}.
\newblock


\bibitem[Tang et~al\mbox{.}(2023)]%
        {tang2023did}
\bibfield{author}{\bibinfo{person}{Ruixiang Tang}, \bibinfo{person}{Qizhang Feng}, \bibinfo{person}{Ninghao Liu}, \bibinfo{person}{Fan Yang}, {and} \bibinfo{person}{Xia Hu}.} \bibinfo{year}{2023}\natexlab{}.
\newblock \showarticletitle{Did you train on my dataset? towards public dataset protection with cleanlabel backdoor watermarking}.
\newblock \bibinfo{journal}{\emph{ACM SIGKDD Explorations Newsletter}} \bibinfo{volume}{25}, \bibinfo{number}{1} (\bibinfo{year}{2023}), \bibinfo{pages}{43--53}.
\newblock


\bibitem[Waheed et~al\mbox{.}(2024)]%
        {waheed2024grove}
\bibfield{author}{\bibinfo{person}{Asim Waheed}, \bibinfo{person}{Vasisht Duddu}, {and} \bibinfo{person}{N Asokan}.} \bibinfo{year}{2024}\natexlab{}.
\newblock \showarticletitle{Grove: Ownership verification of graph neural networks using embeddings}. In \bibinfo{booktitle}{\emph{2024 IEEE Symposium on Security and Privacy (SP)}}. IEEE, \bibinfo{pages}{2460--2477}.
\newblock


\bibitem[Wang et~al\mbox{.}(2004)]%
        {wang2004image}
\bibfield{author}{\bibinfo{person}{Zhou Wang}, \bibinfo{person}{Alan~C Bovik}, \bibinfo{person}{Hamid~R Sheikh}, {and} \bibinfo{person}{Eero~P Simoncelli}.} \bibinfo{year}{2004}\natexlab{}.
\newblock \showarticletitle{Image quality assessment: from error visibility to structural similarity}.
\newblock \bibinfo{journal}{\emph{IEEE transactions on image processing}} \bibinfo{volume}{13}, \bibinfo{number}{4} (\bibinfo{year}{2004}), \bibinfo{pages}{600--612}.
\newblock


\bibitem[Wang et~al\mbox{.}(2021)]%
        {wang2021simvlm}
\bibfield{author}{\bibinfo{person}{Zirui Wang}, \bibinfo{person}{Jiahui Yu}, \bibinfo{person}{Adams~Wei Yu}, \bibinfo{person}{Zihang Dai}, \bibinfo{person}{Yulia Tsvetkov}, {and} \bibinfo{person}{Yuan Cao}.} \bibinfo{year}{2021}\natexlab{}.
\newblock \showarticletitle{Simvlm: Simple visual language model pretraining with weak supervision}.
\newblock \bibinfo{journal}{\emph{arXiv preprint arXiv:2108.10904}} (\bibinfo{year}{2021}).
\newblock


\bibitem[Wen et~al\mbox{.}(2024)]%
        {wen2024hard}
\bibfield{author}{\bibinfo{person}{Yuxin Wen}, \bibinfo{person}{Neel Jain}, \bibinfo{person}{John Kirchenbauer}, \bibinfo{person}{Micah Goldblum}, \bibinfo{person}{Jonas Geiping}, {and} \bibinfo{person}{Tom Goldstein}.} \bibinfo{year}{2024}\natexlab{}.
\newblock \showarticletitle{Hard prompts made easy: Gradient-based discrete optimization for prompt tuning and discovery}.
\newblock \bibinfo{journal}{\emph{Advances in Neural Information Processing Systems}}  \bibinfo{volume}{36} (\bibinfo{year}{2024}).
\newblock


\bibitem[Wen et~al\mbox{.}(2023)]%
        {wen2023enhancing}
\bibfield{author}{\bibinfo{person}{Zhoufutu Wen}, \bibinfo{person}{Xinyu Zhao}, \bibinfo{person}{Zhipeng Jin}, \bibinfo{person}{Yi Yang}, \bibinfo{person}{Wei Jia}, \bibinfo{person}{Xiaodong Chen}, \bibinfo{person}{Shuanglong Li}, {and} \bibinfo{person}{Lin Liu}.} \bibinfo{year}{2023}\natexlab{}.
\newblock \showarticletitle{Enhancing Dynamic Image Advertising with Vision-Language Pre-training}. In \bibinfo{booktitle}{\emph{Proceedings of the 46th International ACM SIGIR Conference on Research and Development in Information Retrieval}}. \bibinfo{pages}{3310--3314}.
\newblock


\bibitem[Xu et~al\mbox{.}(2024)]%
        {xu2024lvlm}
\bibfield{author}{\bibinfo{person}{Peng Xu}, \bibinfo{person}{Wenqi Shao}, \bibinfo{person}{Kaipeng Zhang}, \bibinfo{person}{Peng Gao}, \bibinfo{person}{Shuo Liu}, \bibinfo{person}{Meng Lei}, \bibinfo{person}{Fanqing Meng}, \bibinfo{person}{Siyuan Huang}, \bibinfo{person}{Yu Qiao}, {and} \bibinfo{person}{Ping Luo}.} \bibinfo{year}{2024}\natexlab{}.
\newblock \showarticletitle{Lvlm-ehub: A comprehensive evaluation benchmark for large vision-language models}.
\newblock \bibinfo{journal}{\emph{IEEE Transactions on Pattern Analysis and Machine Intelligence}} (\bibinfo{year}{2024}).
\newblock


\bibitem[Yao et~al\mbox{.}(2023)]%
        {yao2023visual}
\bibfield{author}{\bibinfo{person}{Hantao Yao}, \bibinfo{person}{Rui Zhang}, {and} \bibinfo{person}{Changsheng Xu}.} \bibinfo{year}{2023}\natexlab{}.
\newblock \showarticletitle{Visual-language prompt tuning with knowledge-guided context optimization}. In \bibinfo{booktitle}{\emph{Proceedings of the IEEE/CVF conference on computer vision and pattern recognition}}. \bibinfo{pages}{6757--6767}.
\newblock


\bibitem[Zhang et~al\mbox{.}(2024a)]%
        {zhang2024universal}
\bibfield{author}{\bibinfo{person}{Peng-Fei Zhang}, \bibinfo{person}{Zi Huang}, {and} \bibinfo{person}{Guangdong Bai}.} \bibinfo{year}{2024}\natexlab{a}.
\newblock \showarticletitle{Universal adversarial perturbations for vision-language pre-trained models}. In \bibinfo{booktitle}{\emph{Proceedings of the 47th International ACM SIGIR Conference on Research and Development in Information Retrieval}}. \bibinfo{pages}{862--871}.
\newblock


\bibitem[Zhang et~al\mbox{.}(2024b)]%
        {zhang2024concept}
\bibfield{author}{\bibinfo{person}{Yi Zhang}, \bibinfo{person}{Ce Zhang}, \bibinfo{person}{Ke Yu}, \bibinfo{person}{Yushun Tang}, {and} \bibinfo{person}{Zhihai He}.} \bibinfo{year}{2024}\natexlab{b}.
\newblock \showarticletitle{Concept-Guided Prompt Learning for Generalization in Vision-Language Models}. In \bibinfo{booktitle}{\emph{Proceedings of the AAAI Conference on Artificial Intelligence}}, Vol.~\bibinfo{volume}{38}. \bibinfo{pages}{7377--7386}.
\newblock


\bibitem[Zhao et~al\mbox{.}(2023)]%
        {zhao2023minimizing}
\bibfield{author}{\bibinfo{person}{Anqi Zhao}, \bibinfo{person}{Tong Chu}, \bibinfo{person}{Yahao Liu}, \bibinfo{person}{Wen Li}, \bibinfo{person}{Jingjing Li}, {and} \bibinfo{person}{Lixin Duan}.} \bibinfo{year}{2023}\natexlab{}.
\newblock \showarticletitle{Minimizing maximum model discrepancy for transferable black-box targeted attacks}. In \bibinfo{booktitle}{\emph{Proceedings of the IEEE/CVF conference on computer vision and pattern recognition}}. \bibinfo{pages}{8153--8162}.
\newblock


\bibitem[Zhou et~al\mbox{.}(2022a)]%
        {zhou2022conditional}
\bibfield{author}{\bibinfo{person}{Kaiyang Zhou}, \bibinfo{person}{Jingkang Yang}, \bibinfo{person}{Chen~Change Loy}, {and} \bibinfo{person}{Ziwei Liu}.} \bibinfo{year}{2022}\natexlab{a}.
\newblock \showarticletitle{Conditional prompt learning for vision-language models}. In \bibinfo{booktitle}{\emph{Proceedings of the IEEE/CVF conference on computer vision and pattern recognition}}. \bibinfo{pages}{16816--16825}.
\newblock


\bibitem[Zhou et~al\mbox{.}(2022b)]%
        {zhou2022learning}
\bibfield{author}{\bibinfo{person}{Kaiyang Zhou}, \bibinfo{person}{Jingkang Yang}, \bibinfo{person}{Chen~Change Loy}, {and} \bibinfo{person}{Ziwei Liu}.} \bibinfo{year}{2022}\natexlab{b}.
\newblock \showarticletitle{Learning to prompt for vision-language models}.
\newblock \bibinfo{journal}{\emph{International Journal of Computer Vision}} \bibinfo{volume}{130}, \bibinfo{number}{9} (\bibinfo{year}{2022}), \bibinfo{pages}{2337--2348}.
\newblock


\bibitem[Zhou et~al\mbox{.}(2023)]%
        {AdvCLIP}
\bibfield{author}{\bibinfo{person}{Ziqi Zhou}, \bibinfo{person}{Shengshan Hu}, \bibinfo{person}{Minghui Li}, \bibinfo{person}{Hangtao Zhang}, \bibinfo{person}{Yechao Zhang}, {and} \bibinfo{person}{Hai Jin}.} \bibinfo{year}{2023}\natexlab{}.
\newblock \showarticletitle{Advclip: Downstream-agnostic adversarial examples in multimodal contrastive learning}. In \bibinfo{booktitle}{\emph{Proceedings of the 31st ACM International Conference on Multimedia}}. \bibinfo{pages}{6311--6320}.
\newblock


\bibitem[Zhu et~al\mbox{.}(2023)]%
        {zhu2023prompt}
\bibfield{author}{\bibinfo{person}{Beier Zhu}, \bibinfo{person}{Yulei Niu}, \bibinfo{person}{Yucheng Han}, \bibinfo{person}{Yue Wu}, {and} \bibinfo{person}{Hanwang Zhang}.} \bibinfo{year}{2023}\natexlab{}.
\newblock \showarticletitle{Prompt-aligned gradient for prompt tuning}. In \bibinfo{booktitle}{\emph{Proceedings of the IEEE/CVF International Conference on Computer Vision}}. \bibinfo{pages}{15659--15669}.
\newblock


\bibitem[Zhu et~al\mbox{.}(2024)]%
        {zhu2024reliable}
\bibfield{author}{\bibinfo{person}{Hongyu Zhu}, \bibinfo{person}{Sichu Liang}, \bibinfo{person}{Wentao Hu}, \bibinfo{person}{Li Fangqi}, \bibinfo{person}{Ju Jia}, {and} \bibinfo{person}{Shi-Lin Wang}.} \bibinfo{year}{2024}\natexlab{}.
\newblock \showarticletitle{Reliable Model Watermarking: Defending Against Theft without Compromising on Evasion}. In \bibinfo{booktitle}{\emph{Proceedings of the 32nd ACM International Conference on Multimedia}}. \bibinfo{pages}{10124--10133}.
\newblock


\end{thebibliography}

\end{document}